\documentstyle[rotating,a4wide,epsfig,twoside,12pt]{article}
\renewcommand{\theequation}{\arabic{section}.\arabic{equation}}
\newcommand{\Rotwinkel}{90}

\def\vecgr#1{\mathchoice{\mbox{\boldmath$\mathrm\displaystyle#1$}}
{\mbox{\boldmath$\mathrm\textstyle#1$}}
{\mbox{\boldmath$\mathrm\scriptstyle#1$}}
{\mbox{\boldmath$\mathrm\scriptscriptstyle#1$}}}

\def\vec#1{\mathchoice{\mathrm{\mathbf{\displaystyle#1}}}
{\mathrm{\mathbf{\textstyle#1}}}
{\mathrm{\mathbf{\scriptstyle#1}}}
{\mathrm{\mathbf{\scriptscriptstyle#1}}}}
\newcommand{\be}{\begin{equation}}
\newcommand{\ee}{\end{equation}}
\newcommand{\ba}{\begin{array}}
\newcommand{\ea}{\end{array}}
\newcommand{\bea}{\begin{eqnarray}}
\newcommand{\eea}{\end{eqnarray}}

\newsavebox{\TRS}
\sbox{\TRS}{\hspace{.5em} = \hspace{-1.8em}
                 \raisebox{1ex}{\mbox{\scriptsize TRS}} }

\newsavebox{\defgleich}
\sbox{\defgleich}{\ :=\ }

\newsavebox{\LSIM}
\sbox{\LSIM}{\raisebox{-1ex}{$\ \stackrel{\textstyle<}{\sim}\ $}}

\newsavebox{\GSIM}
\sbox{\GSIM}{\raisebox{-1ex}{$\ \stackrel{\textstyle>}{\sim}\ $}}

\newcommand{\lk}{\left}
\newcommand{\rk}{\right}
\newcounter{saveeqn}

\newcommand{\ssty}{\scriptstyle}
\newcommand{\sssty}{\scriptscriptstyle}
\newcommand{\fns}{\footnotesize}

\newcommand{\ra}{\,\rangle}
\newcommand{\la}{\langle\,}

\newcommand{\ptdnd}[1]{\frac{\partial}{\partial #1}}

\newcommand{\cross}{\times}
\newcommand{\prop}{\propto}

\newcommand{\Tr}{\mbox{Tr$\,$}}

\newcommand{\eins}{\mbox{$1 \hspace{-1.0mm}  {\rm l}$}}
\newcommand{\NatZ}{{\mbox{\,N$\!\!\!\!\!\!\!\:${\rm I}$\,\,\,\,$}}}

\newcommand{\colvecz}[2]{\lk(\ba{c} #1 \\ #2 \ea\rk)}
\newcommand{\colvecd}[3]{\lk(\ba{c} #1 \\ #2 \\ #3 \ea\rk)}

\newcommand{\eh}{\mbox{$\frac{1}{2}$}}

\newcommand{\ev}{\mbox{$\frac{1}{4}$}}

\newcommand{\ih}{\mbox{$\frac{i}{2}$}}
\newcommand{\iv}{\mbox{$\frac{i}{4}$}}

\newcommand{\The}{\Theta}

\newcommand{\alp}{\alpha}
\newcommand{\bet}{\beta}
\newcommand{\gam}{\gamma}
\newcommand{\Gam}{\Gamma}
\newcommand{\del}{\delta}
\newcommand{\Del}{\Delta}

\newcommand{\vep}{\varepsilon}
\newcommand{\kap}{\kappa}
\newcommand{\lam}{\lambda}
\newcommand{\ome}{\omega}

\newcommand{\sig}{\sigma}

\newcommand{\sigmax}{\sigma_{max}}

\newcommand{\brI}{{\mathrm I}}
\newcommand{\brII}{{\mathrm I\!I}}
\newcommand{\rI}{{\mathrm\ssty I}}
\newcommand{\rII}{{\mathrm\ssty I\!I}}

\newcommand{\hpnc}{\mbox{$H_{\sssty\mathrm PNC}$}}

\newcommand{\unheff}{\unH_{e\!f\!\!f}}

\newcommand{\Hw}{H_{\sssty\mathrm w}}

\newcommand{\cB}{{\cal B}}

\newcommand{\cE}{{\cal E}}

\newcommand{\cO}{{\cal O}}

\newcommand{\cR}{{\cal R}}

\newcommand{\cT}{{\cal T}}

\newcommand{\unH}{\mbox{$\underline{H\!}\,$}}

\newcommand{\unU}{\mbox{$\underline{U\!}\,$}}

\newcommand{\unV}{\mbox{$\underline{V\!}\,$}}

\newcommand{\hvh}{{\hat{\hv}}}

\newcommand{\lamtil}{\tilde\lambda}

\newcommand{\ttil}{\tilde t}

\newcommand{\vvtil}{\widetilde{\vv}}
\newcommand{\wtil}{\widetilde{w}}
\newcommand{\wvtil}{\widetilde{\wv}}

\newcommand{\Nbar}{\,\overline{\phantom{{\mathrm N}}}\hspace{-3.5mm}N}

\newcommand{\av}{\vec{a}}

\newcommand{\bv}{\vec{b}}
\newcommand{\Bv}{\vecgr{\cB}}
\newcommand{\BvR}{\Bv_{\mbox{\tiny R}}}

\newcommand{\Dv}{\vec{D}}

\newcommand{\evz}{\vec{e}_2}
\newcommand{\evd}{\vec{e}_3}

\newcommand{\cEv}{\vecgr{\cal E}}

\newcommand{\cEvR}{\cEv_{\mbox{\tiny R}}}

\newcommand{\hv}{\vec{h}}

\newcommand{\sigv}{\vecgr{\sigma}}

\newcommand{\uv}{\vec{u}}

\newcommand{\vv}{\vec{v}}

\newcommand{\wv}{\vec{w}}

\newcommand{\iKsig}{{\sssty (\sigma )}}
\newcommand{\iKtau}{{\sssty (\tau )}}

\newcommand{\iKn}{{\sssty (0)}}
\newcommand{\iKe}{{\sssty (1)}}
\newcommand{\iKz}{{\sssty (2)}}
\newcommand{\iKd}{{\sssty (3)}}

\newcommand{\iKvar}[1]{{\sssty (#1)}}

\newcommand{\lbra}[1]{\mbox{$ ( \widetilde{#1}$}}
\newcommand{\lbrak}[1]{\mbox{$ ( \widetilde{#1}|$}}
\newcommand{\rket}[1]{\mbox{$ | #1 )$}}
\newcommand{\rbra}[1]{\mbox{$ ( #1  $}}
\newcommand{\rbrak}[1]{\mbox{$ ( #1 |$}}

\date{}
\begin{document}
{\sloppy
\begin{titlepage}
%

\title{
{\normalsize 
\hfill \today, HD--THEP--99--21
}
\vspace{1cm}
\\
{\LARGE\bf\sf Large parity violating effects in atomic dysprosium
with nearly degenerate Floquet eigenvalues\thanks{
Work supported by Deutsche Forschungsgemeinschaft, Project No.\ Na 296/1--1} \stepcounter{footnote}}
}

\author{
{\sc 
T. Gasenzer and 
O. Nachtmann}
}

\date{\small\sl 
Institut  f\"ur Theoretische Physik, Universit\"at Heidelberg\\
Philosophenweg 16, D-69120 Heidelberg, Germany
}

\maketitle

\begin{center}
{\bf Abstract:}\\
\parbox[t]{\textwidth}{\small
In this article we study effects of parity nonconservation in atomic dysprosium, where one has a pair of nearly degenerate levels of opposite parity. We consider
the time evolution of this two-level system within oscillatory electric and magnetic fields. These are chosen to have a periodical structure with the same period, such that a Floquet matrix describes the time evolution of the quantum states. We show that, if the states are unstable, the eigenvalues of the Floquet matrix may have contributions proportional to the square root of the parity violating interaction matrix element $\Hw$ while they are almost degenerate in their parity even part. This leads to beat frequencies proportional to $\sqrt{\Hw}$ which are expected to be larger by several orders of magnitude compared to ordinary P-violating contributions which are of order $\Hw$. However, for the simple field configurations we considered, it still seems to be difficult to observe these P-violating beat effects, since the states decay too fast. On the other hand, we found that, within only a few Floquet cycles, very large parity violating asymmetries with respect to experimental setups of opposite chirality may be obtained. The electric and magnetic fields as well as the time intervals necessary for this are in an experimentally accessible range. For statistically significant effects beyond one standard deviation a number of about $10^7$ atoms is required. Our ideas may be applied directly to other 2-level atomic systems and different field configurations. We hope that these ideas will stimulate experimental work in this direction.
} 
\end{center}

\end{titlepage}

%
%
%
%
%
%
%
\section{Introduction}
\label{secIntro}
\setcounter{equation}{0}
In this article we study theoretically possible ways to obtain large parity (P) violating effects in a time-dependent two-level quantum mechanical system. This system is described by a Hamiltonian operator of a fairly general form. However its choice has been motivated by studies of P-violating effects in atomic dysprosium \cite{Budker97}.

Nowadays, using different elements (Cs, Tl, Pb, Bi), atomic parity non-conservation (PNC) measurements have already reached high precisions, for the case of Cs of $\simeq0.3\%$ experimentally, with up to $1\%$ theoretical uncertainty \cite{PNCmeasurements}. (For reviews of PNC in atomic physics cf.\ \cite{Reviews}.)
Dysprosium is another interesting system where atomic PNC can be measured (\cite{Budker97}, \cite{Budker94}, \cite{Budker91}). There is an enhancement of the PNC mixing of a pair of opposite parity levels due to their near degeneracy. The two levels ($4f^{10}\,5d\,6s,\,J=10$), and ($4f^{9}\,5d^2\,6s,\,J=10$) (cf.\ Fig.\ 1 of \cite{Budker97}) are split by an amount of the magnitude of the hyperfine splittings. In particular, the hyperfine components of $^{163}$Dy (nuclear spin: $I=5/2$) with total angular momentum $F=10.5$, the closest pair of opposite parity levels, are split by only $\Del/h= (E_{B}-E_{A})/h=3.1\,$MHz while the separation of them to other levels is at least of the order of GHz (cf.\ Fig.\ 2 of \cite{Budker97}). Moreover with dysprosium one may reduce the experimental uncertainties by comparing the effect for its different stable isotopes.\\

Consider now these states in external electric and magnetic fields $\cEv=\cE\evd$, $\Bv=\cB\evd$, respectively. Here $\evd$ is the unit vector in 3-direction, chosen as the quantization axis, and $\cE$ and $B$ are allowed to vary in time. For suitable values of the magnetic field $\cB$, the sublevels of the opposite parity states with $F=10.5$ and a certain magnetic quantum number, e.g.\ $m_F=-10.5$, are nearly degenerate, i.e.\ the energy separation between them is much smaller than the separation between either of these levels and any other sublevel. Then we may exclusively consider this 2-level subspace $\cR$ with basis
\bea
\label{eq1.1}
  \rket{A} &=& \rket{4f^{10}\,5d\,6s,\,F=10.5,\,m_F=-10.5},\nonumber\\
  \rket{B} &=& \rket{4f^{9}\,5d^2\,6s,\,F=10.5,\,m_F=-10.5}.
\eea
The dynamics in the subspace $\cR$ is governed by an effective Hamiltonian operator, which is represented by a non-hermitean 2$\times$2-matrix 
\be
\label{eq1.2}
  \unheff = \lk(\ba{cc}-\mu_A\cB-\ih\Gam_A  & d\cE+i\,\Hw\\
                       d\cE-i\,\Hw & \Del-\mu_B\cB-\ih\Gam_B \ea\rk),
\ee
where the first row and column refers to level $A$, the second to $B$. The effective Schr\"odinger equation
\be
\label{eq1.3}
  i\ptdnd{t}\,\rket{t} = \unheff\,\rket{t}
\ee
describes the time evolution of the projection $|t)$ of the total time dependent state vector onto the subspace $\cR$ within an approximation where the states $\rket{A}$ and $\rket{B}$ decay exponentially. The corresponding decay widths are $\Gam_{A,B}$. With the standard choice of phases 
for the states (\ref{eq1.1}), the P-violating matrix elements 
\be
\label{eq1.3.1}
  \rbrak{A}\hpnc\rket{B}=-\rbrak{B}\hpnc\rket{A}=i\Hw 
\ee
are purely imaginary due to time reversal (T) invariance. The real constants $d$, $\mu_A$ and $\mu_B$ are related to the electric dipole moment connecting the levels $A$ and $B$ and the magnetic dipole moments respectively: 
\bea
\label{eq1.3.2}
  d(m_F) 
  &:=& -\rbrak{B,F,m_F}D_3\rket{A,F,m_F}
    =  -\rbrak{A,F,m_F}D_3\rket{B,F,m_F}
  \nonumber\\
  &=&  \sum_{m_F}\,\frac{m_J\,\la J,I;m_J,m_I|F,m_F\ra^2}
                        {\sqrt{J(J+1)(2J+1)}} \rbrak{B}|D|\rket{A},\\
\label{eq1.3.3}
  \mu_\alp(m_F)
  &:=& \rbrak{\alp,F,m_F}\mu_3\rket{\alp,F,m_F}\nonumber\\
  &=&  -g_{\alp,F}\,m_F\,\frac{e\hbar}{2m_ec},\qquad(\alp\in\{A,B\}).  
\eea
The $\la J,I;m_J,m_I|F,m_F\ra$ are Clebsch-Gordan coefficients. The reduced matrix element of $\Dv$ was measured to be $|\rbrak{B}|D|\rket{A}|=1.5(1)\cdot10^{-2}e\,a_B$ \cite{Budker94}, where $a_B$ is the Bohr radius and $-e$ and $m_e$ are the charge of the electron and its mass, respectively. The Land\'e factors $g_{\alp,F}$ are given in terms of the factors $g_{\alp,J}$ \cite{Budker94},
%
$g_{\alp,F} = g_{\alp,J}\,[F(F+1)+J(J+1)-I(I+1)]/[2F(F+1)]$,
%
which in turn, are $g_{A,J}=1.21$, $g_{B,J}=1.367$ (cf.\ \cite{Budker94} and \cite{Martin78}).
Numerical values for the constants, which we will use in this paper, are given as follows (cf.\ \cite{Budker97}):
\bea
\label{eq5.1}
  \Gam_A/h       &=& 20\,\mbox{kHz},\nonumber\\
  \Gam_B/h       &=&  1\,\mbox{kHz},\nonumber\\
  d(-10.5)/h     &=&  3.8\,\mbox{kHz V$^{-1}$cm},\nonumber\\
  \Del/h         &=&  3.1\,\mbox{MHz},\nonumber\\
  \mu_A(-10.5)/h &=&  16.4\,\mbox{MHz G$^{-1}$},\nonumber\\
  \mu_B(-10.5)/h &=&  18.5\,\mbox{MHz G$^{-1}$},\nonumber\\
  \Hw/h          &=&  0.002\,\mbox{kHz}.
\eea
Note that in \cite{Budker97}, $\Gam_B/h<1\,$kHz was given.\footnote{
In past experiments, the lifetime of the state showed to be longer than the time of flight of an atom within the apparatus 
\cite{BudkerPrivCom99}.} 
The value for $\Hw$ represents roughly the result for the upper limit $\overline{\Hw}$ on $\Hw$ reported by \cite{Budker97} (cf.\ their Fig.\ 14). We assume this value of $\Hw$ for illustration. Changing to other values of $\Hw$ is straightforward.\\

One common procedure to measure PNC in atomic systems is the following: One prepares an atomic beam with the atoms being in some definite initial state. This state evolves in time as the beam passes through some electromagnetic field configuration. Occupation numbers corresponding to the projection of the resulting final state onto some subspace are measured. If the whole setup has handedness, i.e.\ is non-invariant under a certain improper rotation, PNC effects manifest themselves as changes of such occupation numbers under a reversal of the handedness of the setup.

The observables in such an experiment are thus occupation numbers of states, which should have contributions which change under a P transformation if PNC effects are to be found. For instance, such contributions could be, to leading order, linear in the P violating matrix element $\Hw$. For the above case of dysprosium the time evolution of the states (\ref{eq1.1}) was studied in \cite{Budker97} for a rapidly varying electric field, keeping the magnetic field constant. The frequency of the electric field was chosen large enough such that the changes in the system were non-adiabatic.
From their measurements they report an upper limit on the P-violating matrix element of $|\Hw|/h=|2.4\pm2.9_{stat}\pm0.7_{syst}|\,$Hz \cite{Budker97}. Note that the present theoretical prediction is $\Hw/h=70\pm30\,$Hz \cite{Dzuba94}.\\

The goal of the present paper is to investigate possible enhancement mechanisms for PNC effects in dysprosium. Extensions to other systems are straightforward. For a similar experimental configuration as it as been used in \cite{Budker97} we consider observables, whose sensitivities are large enough to show large P-violating effects and might allow for measurements of the PNC parameters with a higher precision. The underlying ideas have been developed within earlier work on P-violating polarization rotations and P-violating energy shifts in light hydrogen-like atomic systems \cite{BBN95,BGN98,BGN99}. The main new idea now is to study the Floquet operator, i.e.\ the time evolution operator for one cycle of an external potential being periodic in time. It is well known that, knowing the eigenvalues and eigenstates of this operator and the time evolution within {\it one} period, it is 
in general possible 
to calculate the evolution for all times \cite{SchrOp}. In order to simplify our calculations we will consider piecewise constant electric and magnetic fields, which vary discontinuously in time, and use the sudden approximation for the transitions to compute the time evolution of the atomic states.
We will choose the fields such that the $\Hw$-independent parts of the Floquet-eigenvalues are degenerate and investigate the possibility to observe PNC effects which are proportional to the square root of $\Hw$.  The result will be, that, for specific initial states, it is possible to observe, after a definite number of Floquet-periods, very large asymmetries, of order $100\%$, in the occupation number of one of the states $\rket{A},\rket{B}$ w.r.t.\ the reversal of the fields' handedness.\\

The organization of our paper is as follows: In {\it Section \ref{sec2}} we outline the formalism for the description of the time evolution of the atomic states using the Floquet operator. In {\it Section \ref{sec3}} we calculate the eigenvalues and their P-conserving and -violating contributions for a specific electromagnetic field configuration and study the role of the non-hermitecity of the Hamiltonian. In {\it Section \ref{sec4}} we present our PNC observables and consider their statistical significance. Finally, in {\it Section \ref{sec5}}, we present numerical results for the above mentioned states of dysprosium and give our conclusions. Two appendices deal with fundamental aspects concerning the conditions for PNC effects in the Floquet-eigenvalues.

%
%
%
%
%
%
\section{The Floquet operator}
\label{sec2}
\setcounter{equation}{0}
In this section we will derive the Floquet operator which describes the time-evolution of the states of the subspace $\cR$ for the case of external fields $\cE$, $B$ periodic in time, for one oscillation period. The fields are furthermore assumed to be constant during finite time intervals and to change suddenly between them.

Let us assume a periodicity of the fields which sets in at a certain time $t_0\ge0$:
\be
\label{eq2.1}
  \unheff(t_0+nT+\ttil)=\unheff(t_0+\ttil),
\ee
with $n\in\NatZ$, $0\le\ttil<T$. The time evolution of some initial state $\rket{0}$ at time $t=0$ into $\rket{t}$ at time $t>0$ is governed by the Schr\"odinger equation (\ref{eq1.3}), whose solution is given as a time ordered exponential integral:
\be
\label{eq2.2}
  \rket{t} = \cT\exp\lk\{-i\int_0^t\,dt'\,\unheff(t')\rk\}\,\rket{0}.
\ee
With respect to the basis (\ref{eq1.1}) this may be written as a matrix equation,
\be
\label{eq2.3}
  (\alp\rket{t} = \lk(\unU(t,0)\rk)_{\alp\bet}\,(\bet\rket{0}
\ee
($\alp,\bet\in\{A,B\}$; here and in the following we use the summation convention for repeated indices), with
\be
\label{eq2.4}
  \unU(t,t') = \lk((\gam|\cT\exp\lk\{-i\int_{t'}^t\,dt''\,\unheff(t'')\rk\}\,
              \rket{\gam'}\rk),\quad(t\ge t').
\ee
Due to the periodicity of $\unheff$ (\ref{eq2.1}), the evolution matrix $\unU$ is also periodic:
\be
\label{eq2.5}
  \unU(t,t') = \unU(t-T,t'-T),
\ee
for $t_0+T\le t'\le t\le t_0+2T$, and thus
\be
\label{eq2.6}
  \unU(t_0+nT+\ttil,0) = \unU(t_0+\ttil,t_0)\lk[\unU(t_0+T,t_0)\rk]^n\unU(t_0,0)
\ee
($n\in\NatZ$, $0\le\ttil<T$). $\unU(t_0+T,t_0)$ is called the {\sc Floquet} \cite{SchrOp,Floquet} matrix of the periodic Hamilton operator. It becomes immediately clear, that it is only necessary to know the time evolution of the state $\rket{0}$ for $0\le t<t_0+T$ and the {\it eigenvalues} and corresponding {\it eigenvectors} of the Floquet matrix in order to calculate the time evolution up to arbitrary times. Indeed we have 
\be
\label{eq2.7}
  \Big(\unU(t_0+T,t_0)^n\Big)_{\alp\bet} 
  = \sum_{r=\pm} (\alp\rket{r}\,\lam_r^n\,\lbra{r}\rket{\bet},
\ee
where $\lam_\pm$ are the eigenvalues of the (2$\times$2)-matrix $\unU(t_0+T,t_0)$:
\be
\label{eq2.8}
  \lam_\pm = \lam\pm\sqrt{\kap},
\ee
with
\bea
\label{eq2.9}
  \lam &=& \eh\Tr\unU(t_0+T,t_0),\\
\label{eq2.10}
  \kap &=& \lam^2-\det\unU(t_0+T,t_0).
\eea
Furthermore, $\lbra{r}\rket{\bet},(\alp\rket{r}$, $r\in\{+,-\}$, are the components of the left and right eigenvectors of the in general non-hermitean matrix $\unU(t_0+T,t_0)$:
\bea
\label{eq2.11}
  \lk(\unU(t_0+T,t_0)\rk)_{\alp\bet}(\bet\rket{r} 
  &=& \lam_r\,(\alp\rket{r},\\
\label{eq2.12}
  \lbra{r}\rket{\alp}\lk(\unU(t_0+T,t_0)\rk)_{\alp\bet} 
  &=& \lam_r\,\lbra{r}\rket{\bet},
\eea
with the dual normalization
\be
\label{eq2.13}
  \lbra{r}\rket{s} =\del_{rs}
\ee
and the completeness relation in the subspace $\cR$
\be
\label{eq2.14}
  \sum_{r=\pm}\rket{r}\lbrak{r} = \eins,
\ee
and with
\be
\label{eq2.14.1}
  \unU(t_0+T,t_0) = \sum_{r=\pm}\rket{r}\,\lam_r\,\lbrak{r}.
\ee
Let us recall here that the diagonalization of a non-hermitean matrix $\unU$ is always possible if its eigenvalues are all different \cite{Smirnov49}. If there is a degeneracy of eigenvalues one can in general only transform $\unU$ to Jordan's normal form. In our case of (2$\times$2)-matrices this reads
\be
\label{eq2.14.2}
  \lk(\ba{cc}\lam_+&\lam'\\0&\lam_+\ea\rk).
\ee
In the following we will in fact concentrate on the case where the Floquet matrix $\unU(t_0+T,t_0)$ is of the form (\ref{eq2.14.2}), i.e.\ non-diagonalizable without the P-violating terms, but becomes diagonalizable when P-violation is included. This is precisely the case which is needed to get $\sqrt{\Hw}$ P-violating effects. In summary: For the cases which we consider we always have non-degenerate eigenvalues of our Floquet matrix and thus the diagonalization (\ref{eq2.11})--(\ref{eq2.14.1}) is guaranteed. For the points of special interest for us this non-degeneracy is only achieved when P-violation is taken into account.

Let us now specify the configuration of electric and magnetic fields, which we will consider. We assume the fields to point in 3-direction and to be constant within certain time intervals. For this we choose times $t^\iKsig$ ($\sig=1,...,\sigmax$) with
\be
\label{eq2.15}
  0\equiv t^\iKn<t^\iKe<...<t^\iKvar{\sigmax}\equiv T.
\ee
Then, for $0\le\ttil<T$, we write the electric and magnetic fields as
\bea
\label{eq2.16}
  \cEv(t_0+\ttil) &=& \evd\sum_{\sig=1}^{\sigmax}\,\cE^\iKsig
  \The(t^\iKsig-\ttil)\,\The(\ttil-t^\iKvar{\sig-1}),\\
\label{eq2.17}
  \Bv(t_0+\ttil) &=& \evd\sum_{\sig=1}^{\sigmax}\,\cB^\iKsig
  \The(t^\iKsig-\ttil)\,\The(\ttil-t^\iKvar{\sig-1}).
\eea
For arbitrary times $t\ge t_0+T$ these fields are chosen to be
\bea
\label{eq2.18}
  \cEv(t) &=& \cEv(t-T),\\
\label{eq2.19}
  \Bv(t) &=& \Bv(t-T).
\eea
Before $t_0$ they are assumed to be constant:
\bea
\label{eq2.20}
  \cEv(t) &=& \cE^\iKn\evd,\\
\label{eq2.21}
  \Bv(t) &=& \cB^\iKn\evd\quad(0\le t<t_0).
\eea
In Fig.\ \ref{fig2.1} we show an example configuration of $\{\cE^\iKsig,\cB^\iKsig\}$.
%
%
%
%
%
%
%
%
%
%
%
%
\begin{figure}[tb]
\begin{center}
\setlength{\unitlength}{1mm}
\begin{picture}(100,75)(0,-15)
\small
\put(0, 0){\vector(1, 0){105}}
\put(0,-1){\vector(0, 1){30}}
\put(0,35){\vector(1, 0){105}}
\put(0,34){\vector(0, 1){30}}
\multiput(15,0)(30,0){3}{\line(0,-1){2}}
\multiput(24,0)(30,0){3}{\line(0,-1){1}}
\multiput(36,0)(30,0){3}{\line(0,-1){1}}
\multiput(15,35)(30,0){3}{\line(0,-1){2}}
\multiput(24,35)(30,0){3}{\line(0,-1){1}}
\multiput(36,35)(30,0){3}{\line(0,-1){1}}
\put(0,13){\line(1,0){15}}
\put(15,13){\line(0,1){ 6}}
\multiput(15,25)(30,0){3}{\line(1,0){9}}
\multiput(24, 5)(30,0){3}{\line(1,0){12}}
\multiput(36,19)(30,0){2}{\line(1,0){9}}
\multiput(15,25)(30,0){3}{\line(0,-1){ 6}}
\multiput(24, 5)(30,0){3}{\line(0, 1){20}}
\multiput(36,19)(30,0){2}{\line(0,-1){14}}
\put(0,45){\line(1,0){15}}
\put(15,45){\line(0,-1){3}}
\multiput(15,38)(30,0){3}{\line(1,0){9}}
\multiput(24,55)(30,0){3}{\line(1,0){12}}
\multiput(36,42)(30,0){2}{\line(1,0){9}}
\multiput(15,38)(30,0){3}{\line(0, 1){ 4}}
\multiput(24,55)(30,0){3}{\line(0,-1){17}}
\multiput(36,42)(30,0){2}{\line(0, 1){13}}
\put(96,19){$\dots$}
\put(96,42){$\dots$}
\put(103,-5){$t$}
\put(103,30){$t$}
\put(15,-8){\line(1, 0){30}}
\put(15,-8){\line(0,-1){1}}
\put(24,-8){\line(0,-1){1}}
\put(36,-8){\line(0,-1){1}}
\put(45,-8){\line(0,-1){1}}
\put(14,-13){$0$}
\put(23,-13){$t^\iKe$}
\put(35,-13){$t^\iKz$}
\put(44,-13){$t^\iKd\equiv T$}
\put(-1,-6){$0$}
\put(14,-6){$t_0$}
\put(40,-6){$t_0+T$}
\put(70,-6){$t_0+2T$}
\put(-8,15){$\cE^\iKsig$}
\put(-8,50){$\cB^\iKsig$}
\put( 5,32){$\ssty\sig=0$}
\multiput(19,32)(30,0){3}{$\ssty 1$}
\multiput(30,32)(30,0){3}{$\ssty 2$}
\multiput(40,32)(30,0){2}{$\ssty 3$}
\end{picture}
\caption[ ]{\label{fig2.1}\fns Example of the time dependence of the electric and magnetic field strengths. An initial time interval $0<t<t_0$ where the fields are constant is followed by an arbitrary number of Floquet cycles of period $T$. Each Floquet cycle is divided into $\sigmax$ (here: $\sigmax=3$) intervals of length $\Del t^\iKsig=t^\iKsig-t^\iKvar{\sig-1}$.}
\end{center}
\end{figure}
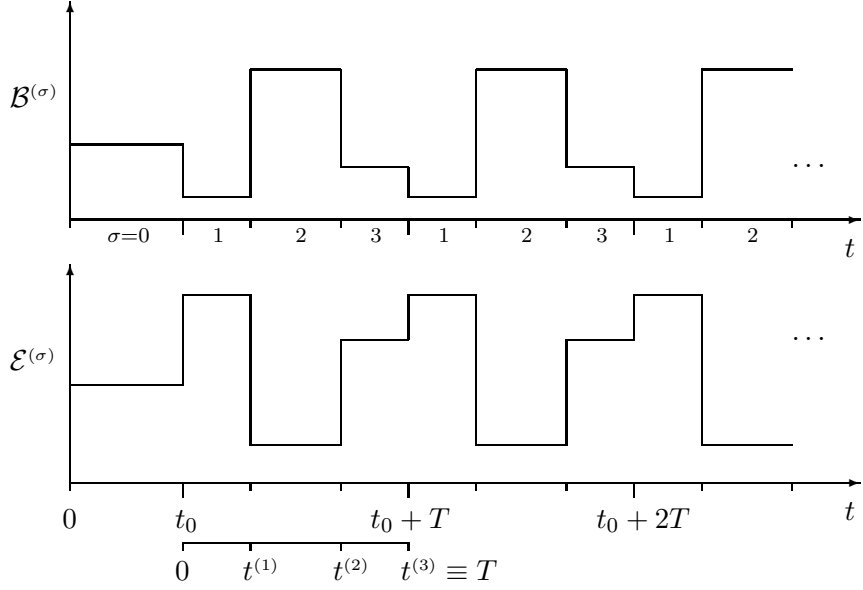
%

%
%

To compute the time evolution of the system, we start with the effective Hamilton operator (\ref{eq1.2}). In the subspace $\cR$, w.r.t.\ the basis (\ref{eq1.1}), it is represented by the (2$\times$2)-matrix
\bea
\label{eq2.22}
  \unheff(t_0+\ttil) 
  &=& \sum_{\sig=0}^{\sigmax}\,\unheff^\iKsig\,
      \The(t^\iKsig-\ttil)\,\The(\ttil-t^\iKvar{\sig-1}),\\
\label{eq2.23}
  \unheff^\iKsig 
  &=& \lk(\ba{cc}-\mu_A\cB^\iKsig-i\Gam_A/2 & d\cE^\iKsig+i\Hw       \\
                 d\cE^\iKsig-i\Hw  & \Del -\mu_B\cB^\iKsig-i\Gam_B/2 \ea\rk),
\eea
where (\ref{eq2.22}) defines $\unheff$ in the range $t_0\le t=t_0+\ttil<t_0+T$. For later times it is defined by the periodicity relation (\ref{eq2.1}). For $0<t<t_0$, it is defined analogously, with the fields (\ref{eq2.20}), (\ref{eq2.21}).

In order to calculate the eigenvalues of the corresponding Floquet matrix we will perform the integration in the exponent of (\ref{eq2.2}) for each time interval with constant fields, where the time ordering plays no role, and then multiply the resulting matrices in the correct order:
\be
\label{eq2.24}
  \unU(t_0+T,t_0) = \cT\prod_{\sig=1}^{\sigmax}\,\unU^\iKsig,
\ee
where
\be
\label{eq2.25}
  \unU^\iKsig = \exp\lk\{-i\unheff^\iKsig(t^\iKsig-t^\iKvar{\sig-1})\rk\}.
\ee
$\cT\prod_\sig$ denotes the multiplication from right to left, with ascending $\sig$. Let us write the segmental (2$\times$2)-matrices $\unheff^\iKsig$ (\ref{eq2.23}) using the SU(2) basis 
\be
\label{eq2.25.1}
   \{\sig_0=\eins_2,\sig_i;i=1,2,3\},
\ee
where the $\sig_i$ are the Pauli spin matrices, as follows:
\be
\label{eq2.26}
  \unheff^\iKsig =: \sum_{i=0}^3\,h_i^\iKsig\sig_i,
\ee
with
\bea
\label{eq2.27}
  h_0^\iKsig &=& \eh\lk[\Del-(\mu_A+\mu_B)\cB^\iKsig-\ih(\Gam_A+\Gam_B)\rk],\\
\label{eq2.28}
  h_1^\iKsig &=& d\,\cE^\iKsig,\\
\label{eq2.29}
  h_2^\iKsig &=& -\Hw ,\\
\label{eq2.30}
  h_3^\iKsig &=& -\eh\lk[\Del+(\mu_A-\mu_B)\cB^\iKsig+\ih(\Gam_A-\Gam_B)\rk].
\eea
Then (\ref{eq2.25}) may be expressed as
\be
\label{eq2.31}
  \unU^\iKsig = \exp\lk\{-ih_0^\iKsig\Del t^\iKsig\rk\}\,
                \sum_{i=0}^3\,u_i^\iKsig\sig_i,
\ee
with
\bea
\label{eq2.32}
  u_0^\iKsig &=& \cos\lk(\eta^\iKsig\Del t^\iKsig\rk),\\
\label{eq2.33}
  u_i^\iKsig &=& -ih_i^\iKsig\,\sin\lk(\eta^\iKsig\Del t^\iKsig\rk)/\eta^\iKsig,
  \ \ i=1,2,3,\\
\label{eq2.34}
  \eta^\iKsig &=& \lk(\sum_{i=1}^3\,{h_i^\iKsig}^2\rk)^{1/2},\\
\label{eq2.35}
  \Del t^\iKsig &=& t^\iKsig-t^\iKvar{\sig-1}.
\eea
Note that the $\eta^\iKsig$ are even functions of $\Hw$.

%
%
%
%
%
%
\section{P-violating contributions to the Floquet eigenvalues}
\label{sec3}
\setcounter{equation}{0}
We would now like to ask the question, under what conditions there may be P-violating contributions to the Floquet eigenvalues $\lam_\pm$ (\ref{eq2.8}). 

To study this consider the field configuration (\ref{eq2.16})--(\ref{eq2.21}) and the one obtained by a reflection R on the 1-2-plane:
\be
\label{eq3.4.1}
  \cEvR^\iKsig = -\cEv^\iKsig,\qquad
   \BvR^\iKsig =   \Bv^\iKsig.
\ee
One signature of P-violation is
\bea
\label{eq3.2}
  \Del\lam_\pm 
  &=& \lam_\pm(\{\cEv^\iKsig,\Bv^\iKsig\};\Hw)
     -\lam_\pm(\{\cEvR^\iKsig,\BvR^\iKsig\};\Hw)\nonumber\\
  &\not=& 0.
\eea
It is easy to see from (\ref{eq2.23}) that we have            
\be
\label{eq3.3}
  \lam_\pm(\{\cEv^\iKsig,\Bv^\iKsig\};\Hw) 
  = \lam_\pm(\{\cEvR^\iKsig,\BvR^\iKsig\};-\Hw)
\ee
and thus
\be
\label{eq3.4}
  \Del\lam_\pm 
  = \lam_\pm(\{\cEv^\iKsig,\Bv^\iKsig\};\Hw)
   -\lam_\pm(\{\cEv^\iKsig,\Bv^\iKsig\};-\Hw).
\ee

Let us now consider the possibility of $\Del\lam_{\pm}\not=0$. From (\ref{eq2.24}), (\ref{eq2.25}), (\ref{eq2.26}) and (\ref{eq2.31}) we find
\be
\label{eq3.5}
  \det\unU(t_0+T,t_0) = \exp\lk\{-2i\sum_\sig\,h_0^\iKsig\Del t^\iKsig\rk\},
\ee
which is thus independent of $\Hw$ (cf.\ (\ref{eq2.27})).
Therefore, taking into account (\ref{eq2.8})--(\ref{eq2.10}), we only have to study the trace of $\unU$, as far as P-violating contributions are concerned. We will first show that for certain combinations of the electric and magnetic field strengths' 3-components $\{\cE^\iKsig,\cB^\iKsig;\sig=1,...,\sigmax\}$, there are no $\Hw$-linear contributions to the trace of the Floquet operator. For this we note that we can insert (2$\times$2)-matrices $\unV$ and $\unV^{-1}$, with $\unV$ non-singular, to the left and right of each $\unU^\iKsig$, respectively. This leaves the trace invariant:
\bea
\label{eq3.9}
  &&\Tr\unU(t_0+T,t_0)\ =\ \Tr\lk(\cT\prod_{\sig=1}^{\sigmax}\,\unU^\iKsig\rk)
  \nonumber\\
  &&\quad=\ \Tr\lk(\unV\unU^\iKvar{\sigmax}\unV^{-1}
                   \unV\unU^\iKvar{\sigmax-1}\unV^{-1}\cdots
                   \unV\unU^\iKvar{1}\unV^{-1}\rk).
\eea
The matrices $\unV$ and $\unV^{-1}$ then act as a transformation rotating the Hamiltonians $\unheff^\iKsig$ in the exponents of $\unU^\iKsig$ (cf.\ (\ref{eq2.25})). 
Consider now the case that the endpoints of all 2-dimensional vectors formed by the field strength components $\cE^\iKsig,\cB^\iKsig$ ($\sig=1,...,\sigmax$) lie on a straight line in the $\cE$-$\cB$-plane, i.e.\ they can be parametrized as follows:
\be
\label{eq3.9.1}
    \colvecz{\cE^\iKsig}{\cB^\iKsig}
  = \colvecz{v_\cE}{v_\cB} 
  + \colvecz{w_\cE}{w_\cB}\,\zeta^\iKsig.
\ee
We show in Appendix \ref{appA} that in this case we can find a matrix $\unV$ which, when applied as in (\ref{eq3.9}), transforms away all $\Hw$-odd terms in the trace of the Floquet operator.\\

Thus we find that the ``points'' ($\cE^\iKsig,\cB^\iKsig$) {\it must not} lie on a straight line in the $\cE$-$\cB$-plane in order to obtain $\Del\lam_\pm\not=0$ (cf.\ (\ref{eq3.4})). This is, of course, only possible for a minimum of three time intervals, i.e.\ $\sigmax\ge3$.\\

We will now provide an expression for $\Tr\unU$ (\ref{eq3.9}), for $\sigmax=3$, in terms of $h_0^\iKsig$ and the vectors $\uv^\iKsig$, $\sig=1,2,3$. Using the relation  
\be
\label{eq3.11}
  (a_0\sig_0+\av\cdot\sigv)(b_0\sig_0+\bv\cdot\sigv)
  = (a_0b_0+\av\cdot\bv)\,\sig_0+(a_0\bv+b_0\av+i[\av\cross\bv])\cdot\sigv
\ee
we find:
\be
\label{eq3.20}
  \unU(t_0+T,t_0) 
  =  \unU^\iKd\unU^\iKz\unU^\iKe 
  =: \sum_{i=0}^3\,u_i(1)\sig_i,
\ee
\bea
\label{eq3.12}
  &&u_0(1) = \eh\Tr\lk(\unU^\iKd\unU^\iKz\unU^\iKe\rk)\nonumber\\
  &&\quad=\ \xi\lk[u_0^\iKd u_0^\iKz u_0^\iKe + (u_0^\iKd\uv^\iKz\cdot\uv^\iKe
         +u_0^\iKz\uv^\iKd\cdot\uv^\iKe +u_0^\iKe\uv^\iKd\cdot\uv^\iKz)\rk.
    \nonumber\\
  &&\quad\quad+\ \lk.i(\uv^\iKd\cross\uv^\iKz)\cdot\uv^\iKe\rk],\\
\label{eq3.12.1}
  &&\uv(1) = \eh\lk\{\unU^\iKd\unU^\iKz\unU^\iKe,\sigv\rk\}\nonumber\\
  &&\quad=\  \xi\lk[
     \uv^\iKd(u_0^\iKz u_0^\iKe+\uv^\iKz\cdot\uv^\iKe)
    +\uv^\iKz(u_0^\iKd u_0^\iKe-\uv^\iKd\cdot\uv^\iKe)
    +\uv^\iKe(u_0^\iKd u_0^\iKz+\uv^\iKd\cdot\uv^\iKz)\rk.\nonumber\\
  &&\quad\quad+\ \lk.
     i\,u_0^\iKd(\uv^\iKz\cross\uv^\iKe)
    +i\,u_0^\iKz(\uv^\iKd\cross\uv^\iKe)
    +i\,u_0^\iKe(\uv^\iKd\cross\uv^\iKz)\rk],
\eea
with
\be
\label{eq3.12.2}
  \xi = \exp\lk\{-i\sum_{\sig=1}^3\,h_0^\iKsig\Del t^\iKsig\rk\}.
\ee
The arguments ``$1$'' refer to the number of Floquet cycles, which will be greater than $1$ later, where we will use again the representation (\ref{eq3.20}) of the time evolution matrix in terms of the Pauli matrices.

Taking into account (\ref{eq2.27})--(\ref{eq2.30}) and (\ref{eq2.32})--(\ref{eq2.35}) we find for the P-even and P-odd contributions $\lam_0$ and $\lam_1$ to $\lam$ (\ref{eq2.9}) the following expressions:
\bea
\label{eq3.12.3}
  \lam &=& \lam_0 + \lam_1 \\
\label{eq3.13}
  \lam_0 &=& \eh\lk[\lam(\Hw)+\lam(-\Hw)\rk] =: \lamtil_0\xi,\\
\label{eq3.15}
  \lam_1 &=& \eh\lk[\lam(\Hw)-\lam(-\Hw)\rk] =: \lamtil_1\xi,
\eea
with
\bea
\label{eq3.14}
  \lamtil_0 &=&      \cos(\eta^\iKd\Del t^\iKd)
                     \cos(\eta^\iKz\Del t^\iKz)
                     \cos(\eta^\iKe\Del t^\iKe)\nonumber\\
  &&\ -\Big[\hv^\iKe\cdot\hv^\iKd
               \frac{\sin(\eta^\iKe\Del t^\iKe)}{\eta^\iKe}
                     \cos(\eta^\iKz\Del t^\iKz)
               \frac{\sin(\eta^\iKd\Del t^\iKd)}{\eta^\iKd}\nonumber\\
  &&\ \phantom{-\Big[}+\mbox{cycl.\ perm.\ in $1,2,3$}\ \Big],\\
\label{eq3.16}
  \lamtil_1 &=& (\hv^\iKe\cross\hv^\iKz)\cdot\hv^\iKd\,
              \frac{\sin(\eta^\iKd\Del t^\iKd)}{\eta^\iKd}
              \frac{\sin(\eta^\iKz\Del t^\iKz)}{\eta^\iKz}
              \frac{\sin(\eta^\iKe\Del t^\iKe)}{\eta^\iKe},
\eea
\be
\label{eq3.16.1}
  (\hv^\iKe\cross\hv^\iKz)\cdot\hv^\iKd 
  = -\frac{1}{2}\,d\,(\mu_A-\mu_B)\,\Hw\,\mbox{$
      \lk|\ba{ccc} 1        & 1        & 1        \\
                   \cE^\iKe & \cE^\iKz & \cE^\iKd \\
                   \cB^\iKe & \cB^\iKz & \cB^\iKd   \ea\rk|$}.
\ee
For the eigenvalues $\lam_\pm$ of the Floquet matrix we now get from (\ref{eq2.8})--(\ref{eq2.10}), (\ref{eq3.5}), (\ref{eq3.13}), and (\ref{eq3.15}):
\bea
\label{eq3.17}
  \lam_\pm 
  &=& \xi\lk(\lamtil_0+\lamtil_1
             \pm\sqrt{\lamtil_0^2-1+2\lamtil_0\lamtil_1+\lamtil_1^2} \rk)
  \nonumber\\
  &=& \lam \pm \sqrt{\kap},\\
\label{eq3.17.1}
  \kap
  &=& \xi^2\lk(\lamtil_0^2-1+2\lamtil_0\lamtil_1+\lamtil_1^2\rk),\\
\label{eq3.17.2}
  \Del\lam_\pm
  &=& \xi\lk\{2\lamtil_1
      \pm\lk[\lk(\lamtil_0^2-1+2\lamtil_0\lamtil_1+\lamtil_1^2\rk)^{1/2}
      \rk.\rk.\nonumber\\
  &&\phantom{\xi\Big\{2\lamtil_1\pm\Big[}-\ \lk.\lk.
             \lk(\lamtil_0^2-1-2\lamtil_0\lamtil_1+\lamtil_1^2\rk)^{1/2}
      \rk]\rk\}.
\eea
We see that in general $\Del\lam_\pm$ will be linear in $\lamtil_1$, i.e.\ linear in $\Hw$.

The basic idea which inspired us to investigate the questions presented in this paper was to explore the possibility of splittings $\Del\lam_\pm$ proportional to the square root of $\Hw$, which, according to (\ref{eq3.17.2}) requires 
\bea
\label{eq3.29}
  \lamtil_0^2 
  &=& 1,\\
\label{eq3.30}
  \lamtil_1 
  &\not=& 0.
\eea
This should yield P-violating effects larger by several orders of magnitude compared to ordinary $\Hw$-linear ones. As we will show this is indeed possible for unstable states $A,B$. It is not possible for the case of stable states  ($\Gam_A=\Gam_B=0$), for which the operators $\unheff^\iKsig$ are hermitean and $\unU(t_0+T,t_0)$ is unitary (cf.\ Appendix \ref{appB}).\\

Next we will write down an expression for the iterated Floquet matrix $\unU(t_0+T,t_0)^n$ (cf.\ (\ref{eq2.6})) in terms of the eigenvalues $\lam_\pm$. Using the left and right eigenvectors (\ref{eq2.11})\,f.\ and their orthonormality (\ref{eq2.13})\,f.\ we find:
\bea
\label{eq3.18}
  \unU(t_0+T,t_0)^n 
  &=& \sum_{r=\pm}\,\rket{r}\,\lam_r^n\,\lbrak{r}\nonumber\\
  &=& \eh(\lam_+^n+\lam_-^n)\eins
     +\eh(\lam_+^n-\lam_-^n)\lk[\,\rket{+}\lbrak{+}-\rket{-}\lbrak{-}\,\rk].
\eea
For $n=1$ this relation reads (cf.\ (\ref{eq2.8}))
\be
\label{eq3.19}
  \unU(t_0+T,t_0) 
  = \lam\eins+\sqrt{\kap}\,\lk[\,\rket{+}\lbrak{+}-\rket{-}\lbrak{-}\,\rk].
\ee
On the other hand we can express $\unU$ in the basis (\ref{eq2.25.1}) as shown in (\ref{eq3.20}).
Comparing with (\ref{eq3.19}) we find
\bea
\label{eq3.21}
  \lam 
  &=& \eh\Tr\unU(t_0+T,t_0) = u_0(1),\\
\label{eq3.22}
  \rket{+}\lbrak{+}-\rket{-}\lbrak{-}
  &=& \frac{1}{\sqrt{\kap}}\sum_{i=1}^3\,u_i(1)\sig_i.
\eea
From (\ref{eq3.17}) we see that we can write $\lam_\pm$ as
\bea
\label{eq3.23}
  \lam_\pm &=& \xi\,\exp(\pm\vep),\nonumber\\
  \tanh\vep
  &=& \lk[1-\lk(\lamtil_0+\lamtil_1\rk)^{-2}\rk]^{1/2}\nonumber\\
  &=& \frac{\sqrt{\kap}}{\lam},
\eea
where $\vep$ is in general complex. This gives
\bea
\label{eq3.25}
  \lam_\pm^n 
  &=& \xi^n\exp\lk(\pm n\vep\rk) = \exp\lk(-i\ome_\pm nT\rk),\nonumber\\
  \ome_\pm
  &=&i\lk(\ln\xi\pm\vep\rk)\,T^{-1}.
\eea
Thus we have
\be
\label{eq3.26}
  \unU(t_0+T,t_0)^n = \sum_{i=0}^3\,u_i(n)\sig_i,
\ee
with
\bea
\label{eq3.27}
  u_0(n) 
  &=& \xi^n\cosh\lk(n\vep\rk),\\
\label{eq3.28}
  u_i(n) 
  &=& u_i(1)\xi^n\sinh\lk(n\vep\rk)/\sqrt{\kap},\quad i=1,2,3.
\eea
From (\ref{eq3.26})--(\ref{eq3.28}) we see that for nearly degenerate eigenvalues ($|\kap|\ll1$), there are oscillations of the matrix elements of the Floquet matrix where the (complex) frequencies $\ome_\pm$ contain terms proportional to $\sqrt{\kap}$. For fixed $n$ and $\kap\rightarrow0$, however, there are clearly no terms $\prop\sqrt{\kap}$ in $u_i(n)$ (\ref{eq3.28}), as we see from (\ref{eq3.23}). This is crucial for consistency: Consider the case 
(\ref{eq3.29})\,f.,
which we intend to study in the following sections. Then the oscillation frequencies become proportional to $\sqrt{\kap}\simeq\sqrt{2\lam_1}\prop\sqrt{\Hw}$. But for fixed $n$, no odd powers of $\sqrt{\Hw}$ can appear in an expansion of $\unU$ w.r.t.\ $\sqrt{\Hw}$ since the Hamiltonian operator $\unheff$ contains only a term linear in $\Hw$. 

We emphasize that $\sqrt{\Hw}$ effects in the Floquet frequencies are only possible for unstable states. In Appendix \ref{appB} we show that for stable states, i.e. for $\Gam_A=\Gam_B=0$ frequencies proportional to $\sqrt{\Hw}$ are excluded.

%
%
%
%
%
%
\section{P-violating observables}
\label{sec4}
\setcounter{equation}{0}
In this section we will study possibilities to obtain optimized observables to measure the P-violating contribution $\Hw$ to the Hamiltonian operator for systems of the form described in section \ref{sec2}. The general idea is to measure occupation numbers of the states $\rket{A}$ and $\rket{B}$ of the subspace $\cR$ after a given number $n$ of oscillations of the external fields. In other words, we will consider the moduli squared of some linear combination of the matrix elements of the evolution matrix $\unU(t_0+nT,t_0)$. The initial state $\rket{0}$ is chosen to be either $\rket{A}$ or $\rket{B}$, and the fields $\cE^\iKn\evd$, $\cB^\iKn\evd$, switched on between $t=0$ and $t=t_0$ are used to provide a mixing of these states before the atoms enter the oscillatory field. In the following we will label this experimental setup as roman no.\ $\brI$. At $t=t_0+nT$ the population of either $\rket{A}$ or $\rket{B}$ shall be measured. A measurement of such a population number for a sequence of $n$-values will reveal the Floquet frequencies $\ome_\pm$ (\ref{eq3.25}).

%
%
Suppose we start with a state $\rket{\bet}$ ($\bet=A$ or $B$) at time $t=0$. We let this state evolve until a time $t_0+nT$, when the state vector of the undecayed atom is 
\be
\label{eq4.1}
  \rket{\bet,n;\brI} = U(t_0+nT,0;\{\cEv^\iKsig, \Bv^\iKsig \})\,\rket{\bet}.
\ee
Its norm and thus the probability to find the atom undecayed is
\be
\label{eq4.2}
  w_{\rI}(\bet,n) = \rbra{\bet,n;\brI}\rket{\bet,n;\brI}.
\ee
The probability to find the atom in the state $\alp$ ($\alp\in\{A,B\}$) is
\be
\label{eq4.2.1}
  w_{\rI}(\alp,\bet,n) = |\rbra{\alp}\rket{\bet,n;\brI}|^2.
\ee
We also consider the evolution of the state $\rket{\bet}$ in the reflected setup with
the R-transformed fields $\cEvR^\iKsig,\BvR^\iKsig$ (\ref{eq3.4.1}). The corresponding states and other quantities will be labelled with a roman number $\brII$ as opposed to those corresponding to the original setup, labelled by $\brI$: $\rket{\bet,n;\brII}$, $w_{\rII}(\bet,n)$, etc.

Suppose now that we repeat the experiment with the $\{\cEv^\iKsig, \Bv^\iKsig \}$ setup $N_{0,\rI}$ times starting always at time $t=0$ with the atom in state $\rket{\bet}$. The second experiment with the R-reflected setup $\{\cEvR^\iKsig, \BvR^\iKsig \}$ we repeat $N_{0,\rII}$ times. Let $N_{\rI,\rII}(\bet,n)$ and $N_{\rI,\rII}(\alp,\bet,n)$ be the number of undecayed and the number of atoms in state $\alp$ ($\alp\in\{A,B\}$), respectively, at time $t_0+nT$.
 
Let us assume that we choose the numbers of repetitions $N_{0,\rI}$, $N_{0,\rII}$ such that the numbers of undecayed atoms at time $t_0+nT$ are the same for each setup:
\be
\label{eq4.5}
   N_{\rI}(\bet,n) = N_{\rII}(\bet,n) \equiv N(\bet,n).
\ee
The relative occupation of $\rket{\alp}$ compared to the number of undecayed states, $N_{\rI,\rII}(\alp,\bet,n)/N_{\rI,\rII}(\bet,n)$, is our observable. Its mean value is
\be
\label{eq4.7}
   r_{\rI,\rII}(\alp,\bet,n) 
   = \frac{\la N_{\rI,\rII}(\alp,\bet,n)\ra}{N_{\rI,\rII}(\bet,n)}
   = \frac{w_{\rI,\rII}(\alp,\bet,n)}{w_{\rI,\rII}(\bet,n)}.
\ee
The expected numbers of repetitions necessary to achieve $N(\bet,n)$ undecayed atoms are
\be
\label{eq4.7.1}
   \la N_{0,{\rI,\rII}} \ra 
   = \frac{N_{\rI,\rII}(\bet,n)}{w_{\rI,\rII}(\bet,n)}.
\ee
For an ensemble of $N(\bet,n)$ undecayed atoms the number of atoms in state $\rket{\alp}$ has a variance 
\bea
\label{eq4.8}
   \Big(\Del N_{\rI,\rII}(\alp,\bet,n)\Big)^2 
   &=& \Big(N(\bet,n)\,\Del r_{\rI,\rII}(\alp,\bet,n)\Big)^2\nonumber\\
   &=& N(\bet,n)\,
       r_{\rI,\rII}(\alp,\bet,n)\Big(1-r_{\rI,\rII}(\alp,\bet,n)\Big).
\eea
Suitable P-odd observables are the asymmetries between the relative occupation numbers for the original and the reflected setups:
\be
\label{eq4.9}
  A(\alp,\bet,n)
  = \frac{r_{\rI}(\alp,\bet,n)-r_{\rII}(\alp,\bet,n)}
         {r_{\rI}(\alp,\bet,n)+r_{\rII}(\alp,\bet,n)}.
\ee
Let us now consider the statistical significance of the P-violating observable (\ref{eq4.9}). The variance of (\ref{eq4.9}) for an ensemble of $N(\bet,n)$ undecayed atoms at time $t_0+nT$ is given by 
\bea
\label{eq4.9.1}
  \lk(\Del A(\alp,\bet,n)\rk)^2
  &=& 4\,\frac{r_{\rII}^2\Del r_{\rI}^2 + r_{\rI}^2\Del r_{\rII}^2}
            {(r_{\rI} + r_{\rII})^4}\nonumber\\
  &=& \frac{4}{N(\bet,n)}\,
      \frac{r_{\rI}r_{\rII}(r_{\rI}+r_{\rII}-2\,r_{\rI}r_{\rII})}
            {(r_{\rI} + r_{\rII})^4}\nonumber\\
  (r_{\rI,\rII} &\equiv& r_{\rI,\rII}(\alp,\bet,n)).
\eea
We can claim to see a $1$-$\sig$-effect if 
\be
\label{eq4.9.2}
  \lk|A(\alp,\bet,n)\rk|^2
  > \lk(\Del A(\alp,\bet,n)\rk)^2.
\ee
Using (\ref{eq4.7.1}) this can be written as a condition on the mean total number of initial atoms in the two experiments with opposite chiralities of the field configuration which are necessary to establish a 1-$\sig$-effect: 
\bea
\label{eq4.9.3}
  \la N_0\ra 
  &=& \la N_{0,\rI}\ra + \la N_{0,\rII}\ra > \Nbar_0,\\
\label{eq4.9.4}
  \Nbar_0 
  &=& \lk(\frac{1}{w_{\rI}(\bet,n)}+\frac{1}{w_{\rII}(\bet,n)}\rk)\,
      \frac{4\,r_{\rI}r_{\rII}(r_{\rI}+r_{\rII}-2\,r_{\rI}r_{\rII})}{(r_{\rI} 
      + r_{\rII})^2(r_{\rI} - r_{\rII})^2}.
\eea
The task will now be to minimize the value of $\Nbar_0$ as a function of $n$ and the field strengths $\{\cEv^\iKsig,\Bv^\iKsig;\sig=0,...,\sigmax\}$.

%
%
%
%
%
\section{Results}
\label{sec5}
\setcounter{equation}{0}
In this section we present our numerical results. In the first place we calculate a solution for the fields $\{\cEv^\iKsig, \Bv^\iKsig; \sig=1,2,3 \}$ for which the eigenvalues (\ref{eq3.17}) of the Floquet matrix satisfy the conditions (\ref{eq3.29})\,f., i.e.\ are only split by a P-violating quantity proportional to $\sqrt{\Hw}$. This corresponds to $\lamtil_0^2=1$. Using this solution we calculate the asymmetry (\ref{eq4.1}) and minimize the number $\Nbar_0$ (\ref{eq4.9.4}) of initial atoms required for a 1-$\sig$-effect, by choosing a convenient initial mixing through $\cEv^\iKn,\Bv^\iKn$.
For our calculations we used the fixed constants (cf.\ \cite{Budker97}) as given in (\ref{eq5.1}).

%
%
%
%
%
%
%
%
%
%
%
%
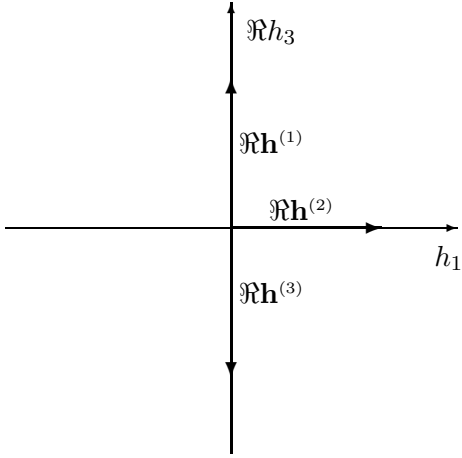
\begin{figure}[tb]
\begin{center}
\setlength{\unitlength}{1mm}
\begin{picture}(100,70)(0,0)
\small
\put(20,35){\vector(1, 0){60}}
\put(50, 5){\vector(0, 1){60}}
\thicklines
\put(50,35){\vector(0, 1){20}}
\put(50,35){\vector(0,-1){20}}
\put(50,35){\vector(1, 0){20}}
%
%
\put(77,30){$h_1$}
\put(52,60){$\Re h_3$}
\put(51,45){$\Re\hv^\iKe$}
\put(55,36){$\Re\hv^\iKz$}
\put(51,25){$\Re\hv^\iKd$}
\end{picture}
\caption[ ]{\label{fig5.1}\fns Sketch of the real parts of the vectors $\hv^\iKsig$ for the choice given in Table \ref{tab5.1}.}
\end{center}
\end{figure}
%

%
%
%
%
%
%
%
%
%
%
%
%
%
\small

\begin{table}[tb]
\renewcommand{\arraystretch}{1.5}
\renewcommand{\tabcolsep}{0.6cm}
\begin{center}

\begin{tabular}{c|ccc}
\hline
\hline 
 $\sig$ & $1$ & $2$ & $3$\\
\hline
 $d\cE^\iKsig/h$ [kHz] & $-8.47\cdot10^{-2}$ & $157.0$  & $-8.47\cdot10^{-2}$\\
 $\cE^\iKsig$ [V/cm]   & $-2.23\cdot10^{-2}$ & $41.3$   & $-2.23\cdot10^{-2}$\\
\hline
 $\eh[\Del+(\mu_A-\mu_B)\cB^\iKsig]/h$ [kHz] 
                & $157.1$ & $0$    & $-157.1$\\
 $\cB^\iKsig$ [G] & $1.33$  & $1.48$ & $1.63$  \\
\hline 
 $\Del t^\iKsig$ [$\mu$s]     & $1.59$ & $9.91\cdot10^{-2}$ & $1.59$\\
\hline 
\hline
\end{tabular}
\caption[ ]{\label{tab5.1}\fns Configuration of the electromagnetic field within the sections $\sig=1,2,3$ of one Floquet cycle (cf.\ (\ref{eq2.6})\,f., (\ref{eq2.35})). At least one of the parameters has to be fine-tuned to more decimals than indicated in order to match exactly the degeneracy condition (\ref{eq3.17.1}). The intervals $\Del t^\iKsig$, $\sig=1,3$, correspond to $2\pi\Del t^\iKsig=0.01\,$(kHz)$^{-1}$.}
\end{center}
\renewcommand{\arraystretch}{1.0}
\end{table}
\normalsize

%
%
%
%
\subsection{$\sqrt{\Hw}$-linear eigenvalues}
\label{sec5.1}
As can be seen from (\ref{eq3.14}), (\ref{eq3.16}), the conditions (\ref{eq3.29})\,f.\ may only be fulfilled for a convenient relative adjustment of the three vectors $\hv^\iKsig$, $\sig=1,2,3$ defined in (\ref{eq2.26})--(\ref{eq2.30}).  

However, we will start by choosing appropriate time intervals $\Del t^\iKsig$ (\ref{eq2.35}). As it is clear from (\ref{eq3.26})\,ff., (\ref{eq4.1}) both the probabilities $w_\pm(\bet,n)$ (\ref{eq4.2}) and the minimum initial atom number $\Nbar_0$ (\ref{eq4.9.4}) are governed by the overall factor $\xi$, viz.\ are proportional to $|\xi|^{2n}$ and $|\xi|^{-2n}$ respectively. From (\ref{eq2.15}), (\ref{eq2.27}), (\ref{eq3.12.1}) we find:
\bea
\label{eq5.2}
  |\xi|^{2n}    &=& \exp\lk\{-\eh(\Gam_A+\Gam_B)T n\rk\},\\
\label{eq5.3}
  T             &=& \sum_{\sig=1}^3\,\Del t^\iKsig.
\eea
%
%
%
%
%
%
%
%
%
%
%
%
%
\begin{figure}[tb]
\begin{center}
\setlength{\unitlength}{0.7mm}

\begin{picture}(190,110)(0,-15)

\put(0,0){\epsfig{file=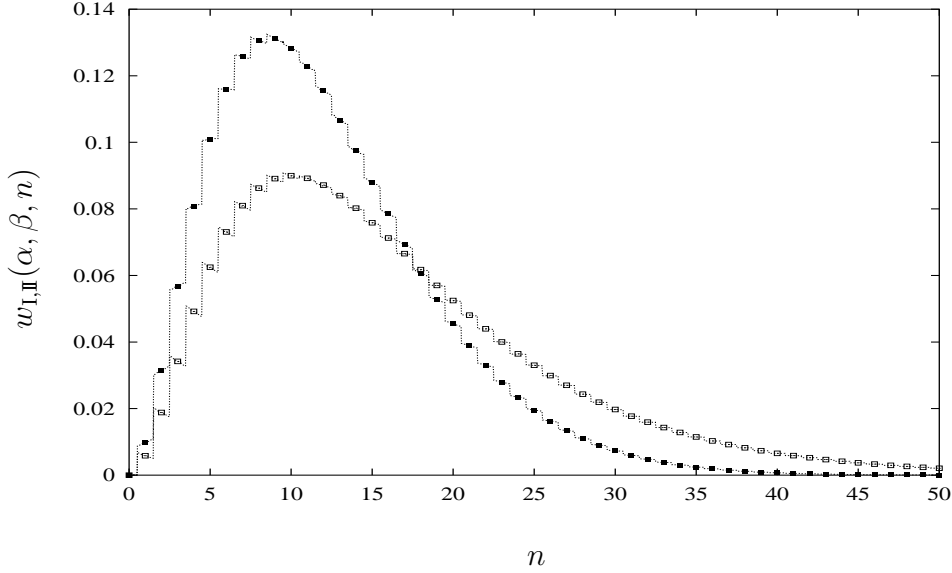,width=180\unitlength,height=100\unitlength}}


\put( 94,-10){$n$}

\begin{rotate}{\Rotwinkel}
\put(35,0){$w_{\rI,\rII}(\alp,\bet,n)$}
\end{rotate}

\end{picture}
\caption[ ]{\label{fig5.7}\fns The probabilities $w_{\rI,\rII}(\alp,\bet,n)$ (\ref{eq4.2.1}) for $\alp=B$, $\bet=A$, $\Hw/h=1\,$kHz, as a function of the number of Floquet cycles $n$. The electric and magnetic fields are chosen as in Table \ref{tab5.1}. No premixing is chosen: $t_0=0$. The open boxes indicate the values of $w_{\rI}(\alp,\bet,n)$, the filled boxes $w_{\rII}(\alp,\bet,n)$, i.e.\ those for the R-reflected setup. The dashed lines represent the complete time evolution of $|\rbrak{\alp}\unU(t,0)\rket{\bet}|^2$ as a function of $t$ which is given in units of $T$. This plot illustrates the first Floquet beat for an hypothetical value for $\Hw$, where the overall decay does not yet overwhelm the P-violating effect in the first beat.}
\end{center}
\end{figure}

%
%
%
%
%
%
%
%
%
%
%
%
%
\begin{figure}[tb]
\begin{center}
\setlength{\unitlength}{0.7mm}

\begin{picture}(190,110)(0,-15)

\put(0,0){\epsfig{file=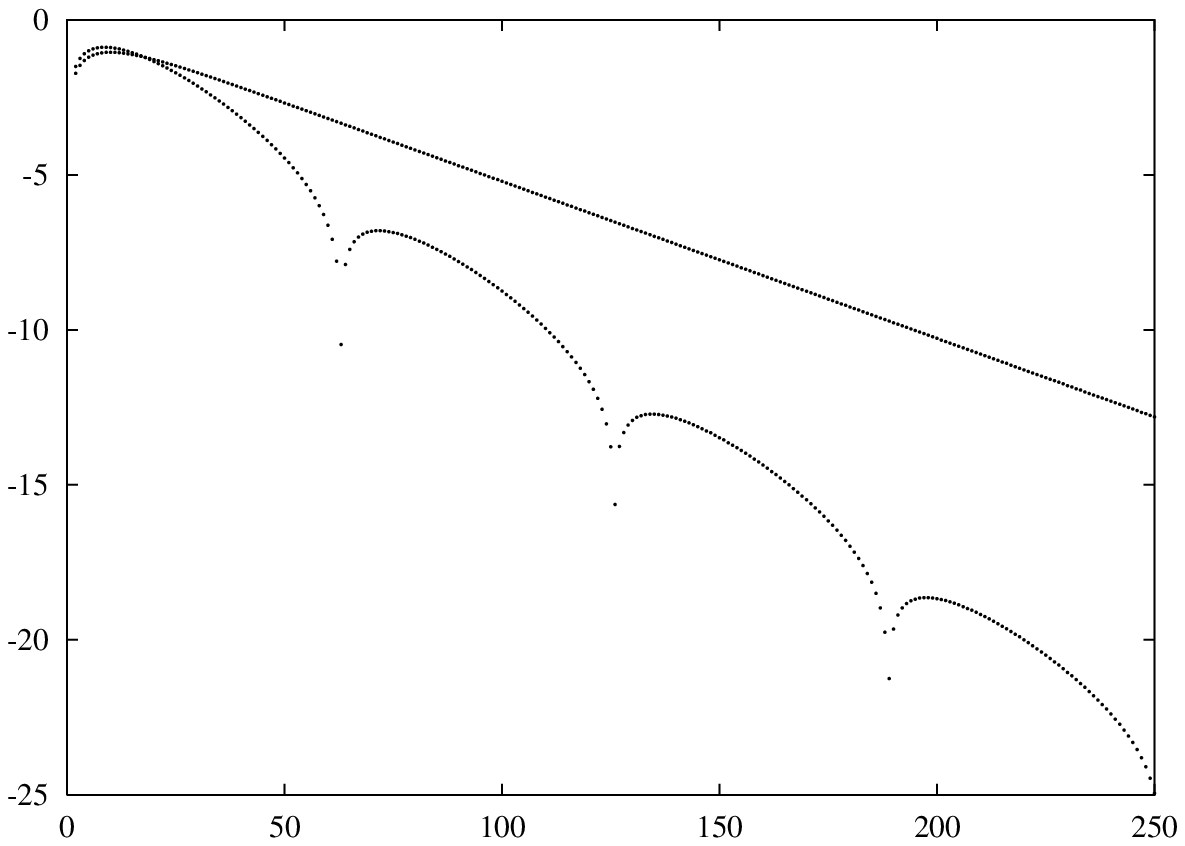,width=180\unitlength,height=100\unitlength}}


\put( 94,-10){$n$}

\begin{rotate}{\Rotwinkel}
\put(30,0){$\lg(w_{\rI,\rII}(\alp,\bet,n))$}
\end{rotate}

\end{picture}
\caption[ ]{\label{fig5.8}\fns The logarithm of the probabilities $w_{\rI,\rII}(\alp,\bet,n)$ (\ref{eq4.2.1}) for the same configuration as in Fig.\ \ref{fig5.7}, as a function of the number of Floquet cycles $n$. Compared to Fig.\ \ref{fig5.7}, the first four Floquet cycles are shown. The oscillatory, i.e.\ the lower sequence of points refers to $w_{\rII}(\alp,\bet,n)$, the upper one to $w_{\rI}(\alp,\bet,n)$.}
\end{center}
\end{figure}

%
%
%
In order to preserve the occupation numbers from being essentially zero before one period of the time evolution within the oscillatory fields, one has to choose the $\Del t^\iKsig$ small enough. Stated differently, we would like to study the case, where a P-violating effect is accumulated during several up to many repeated cycles of the periodic fields. In this case the occupation numbers should show a P-violating oscillation with a wave number proportional to the imaginary part of $\sqrt{\kap}$.
As one derives from the numerical values given in (\ref{eq5.1}), a Floquet time $T$ (\ref{eq5.3}) of
\be
\label{eq5.4}
 T = 3.5\,\mu\mbox{s}
\ee
yields a decay by a factor of $|\xi|^{2n}\simeq 1/10$ within $n=10$ periods.

In order to have a non-vanishing $\lamtil_1$ (\ref{eq3.16}) one has to choose the fields and time intervals $\Del t^\iKsig$ in a way, that the sines in (\ref{eq3.16}) are non-zero. On the other hand we would like $\lamtil_0$ to equal plus or minus 1. As one finds numerically, for the total sum of the time intervals $\Del t^\iKsig$, $\sig=1,2,3$, not to exceed the value (\ref{eq5.4}), this is only possible, if the sines as well as the cosine in the second term of (\ref{eq3.14}) are close to $+1$ or $-1$. (Then the two other terms of the sum over the cyclic permutations in $1,2,3$ in (\ref{eq3.14}) are close to zero.) A different choice, where more than one cosine $\cos(\eta^\iKsig\Del t^\iKsig)$ is away from zero, leads to a reduced, non-optimal value of the modulus of $\lamtil_1$. Now, if two of the sines are to be close to 1 and (\ref{eq5.4}) is supposed to be the upper limit of (\ref{eq5.3}), the electric and magnetic field terms $d\cE^\iKsig$, $[\Del+(\mu_A-\mu_B)\cB^\iKsig]/2$ in the corresponding two segments have to be chosen large compared to the imaginary part $(\Gam_A-\Gam_B)/4$ of $h_3^\iKsig$ (cf.\ (\ref{eq2.28}), (\ref{eq2.30}), (\ref{eq2.34})).

We have chosen a configuration as given in Table \ref{tab5.1}. The representation of the projection of the real parts of the vectors $\hv^\iKsig$ is moreover sketched in Fig.\ \ref{fig5.1}. From this it becomes clear, that the dot product $\hvh^\iKe\cdot\hvh^\iKd$
of the ``unit'' vectors $\hvh^\iKsig:=\hv^\iKsig/\eta^\iKsig$ in the relevant term of (\ref{eq3.14}) is close to $-1$ (only close, because the imaginary parts of the $h_3^\iKsig$ introduce small perturbations). Similarly we have $\hvh^\iKe\cdot\hvh^\iKz\simeq0$, $\hvh^\iKz\cdot\hvh^\iKd\simeq0$. Then we get for $\lamtil_0$ from (\ref{eq3.14}):
\bea
\label{eq5.4.1}
 \lamtil_0
 &\simeq& \cos(\eta^\iKd\Del t^\iKd)
          \cos(\eta^\iKz\Del t^\iKz)
          \cos(\eta^\iKe\Del t^\iKe)\nonumber\\
 & &\ +\  \sin(\eta^\iKd\Del t^\iKd)
          \cos(\eta^\iKz\Del t^\iKz)
          \sin(\eta^\iKe\Del t^\iKe).
\eea
Now we make the factor $\cos(\eta^\iKz\Del t^\iKz)$ to be close to $1$ by choosing the time interval $\Del t^\iKz$ small compared to the two other ones. Moreover we choose $\eta^\iKd\Del t^\iKd\simeq\eta^\iKe\Del t^\iKe$ (cf.\ Table \ref{tab5.1}). In this way we clearly have $\lamtil_0\simeq1$ and by fine tuning of the parameters we can achieve $\lamtil_0=1$. This choice of parameters yields that $\sin(\eta^\iKz\Del t^\iKz)/\eta^\iKz\simeq\Del t^\iKz$ and hence that $\lamtil_1$ is proportional to $\Del t^\iKz$. We have not found a way to avoid such a relatively small proportionality factor, since the alternative choice of a small $\eta^\iKz$ would result in an equivalent reduction in the triple product of the vectors $\hv^\iKsig$ in $\lamtil_1$, which is equal to the area of the triangle spanned by the projections of these vectors on to the 1--3-plane (cf.\ Fig.\ \ref{fig5.1}).

%
%
%
%
%
%
%
%
%
%
%
\begin{figure}[tb]
\begin{center}
\setlength{\unitlength}{0.7mm}

\begin{picture}(190,110)(0,-15)

\put(0,0){\epsfig{file=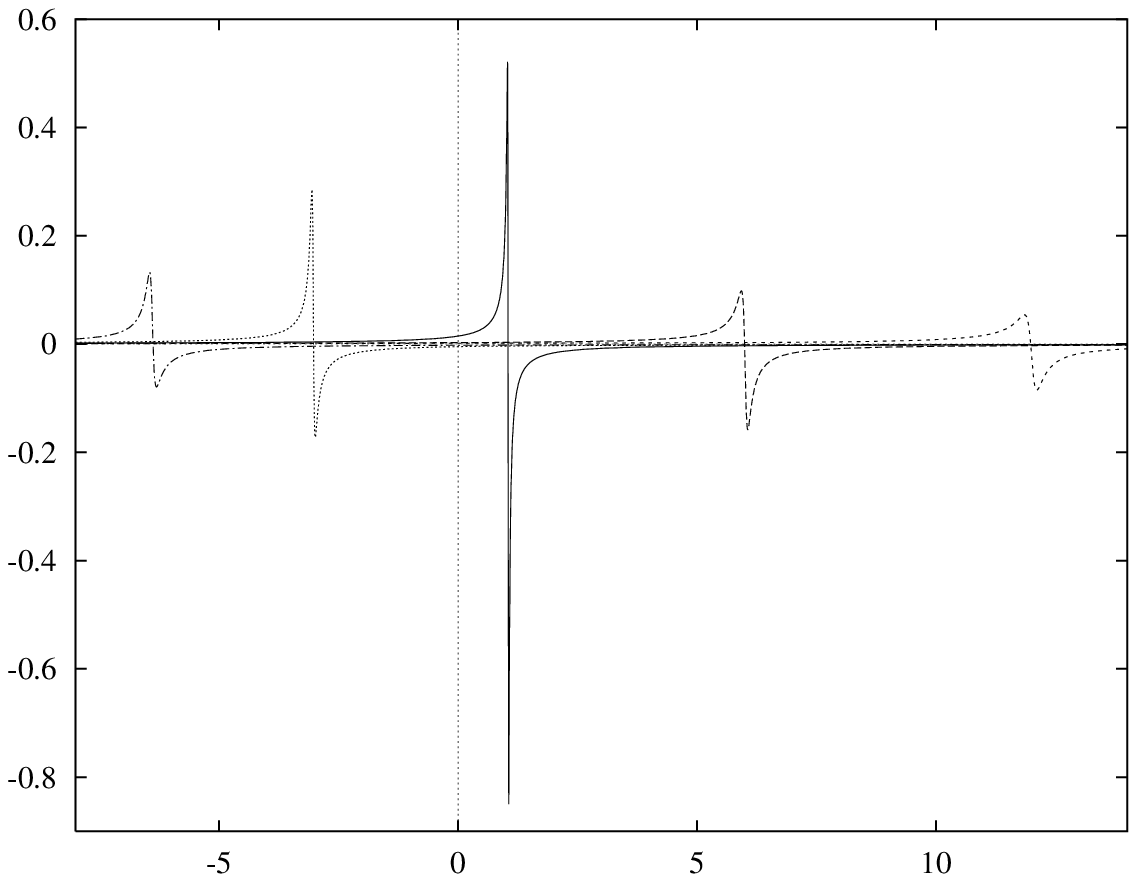,width=180\unitlength,height=100\unitlength}}


\put( 80,-10){$d\cE^\iKn/h$ [kHz]}

\begin{rotate}{\Rotwinkel}
\put(40,-5){$A(\alp,\bet,n)$}
\end{rotate}

\put(28,71){$\ssty n\,=\,12$}
\put(57,71){$\ssty 11$}
\put(85,71){$\ssty 10$}
\put(117,71){$\ssty 9$}
\put(158,71){$\ssty 8$}

\end{picture}
\caption[ ]{\label{fig5.2}\fns The asymmetry $A(\alp,\bet,n)$ (\ref{eq4.9}) for $\alp=\bet=A$, $8\le n\le12$, $\{\cEv^\iKsig,\Bv^\iKsig,\Del t^\iKsig; \sig=1,2,3\}$ as given in Tab.\ \ref{tab5.1}, $\Del+(\mu_A-\mu_B)\cB^\iKn=0$, as a function of $d\cE^\iKn$. Contributions of order $\Hw^2$ have been neglected. The curves are essentially zero except for the small intervals, where resonances occur. The approximate centres of these intervals are given in Table \ref{tab5.2}. Each resonance represents a different $n$. Hence, the largest one at $d\cE^\iKn/h\simeq1\,$kHz corresponds to $n=10$.}
\end{center}
\end{figure}

%
%
In summary, using the values given in Tab.\ \ref{tab5.1}, we obtain:
\bea
\label{eq5.5}
  \lamtil_0 
  &=& 1.0,\\
\label{eq5.6}
  \lamtil_1 
  &=& 1.25\cdot10^{-3}\,(\mbox{kHz}\,h)^{-1}\,\Hw \nonumber\\
  &=& 2.49\cdot10^{-6},\\
\label{eq5.7}
  2\vep
  &=& 4.46\cdot10^{-3},\\
\label{eq5.7.1}
  \ome_\pm(\{\cEv^\iKsig,\Bv^\iKsig\})/h
  &=& -i\,(5.25 \mp 0.108)\,\mbox{kHz},\\
\label{eq5.7.2}
  \ome_\pm(\{\cEvR^\iKsig,\BvR^\iKsig\})/h
  &=& -i\,(5.25 \mp i\,0.108)\,\mbox{kHz},\\
\label{eq5.8}
  \eh(\Gam_A+\Gam_B)T
  &=& 2.17\cdot10^{-1}.
\eea
(Here $\lamtil_1$ is real and positive by chance, which is not true for a general choice of the field configuration.)
As we have mentioned before the fraction of the P-violating oscillation wave number (\ref{eq5.7}) divided by the total decay width (\ref{eq5.8}) is 2\%: 
\be
\label{eq5.8.1}
  \frac{4\,\vep}{(\Gam_A+\Gam_B)T}
  = 2.06\cdot10^{-2}.
\ee
Our numerical investigations showed that this cannot be increased by more than a few tenths of a percent. It is essentially governed by the fraction $\sqrt{\Hw}/(\Gam_A+\Gam_B)$. Longer lived states with smaller decay widths will not a priori lead to a larger fraction (\ref{eq5.8.1}), since then $\Del t^\iKz$ has to be smaller in order to fulfill (\ref{eq5.5}). 

For illustration we show in Figs.\ \ref{fig5.7}, \ref{fig5.8} the probabilities $w_{\rI,\rII}(\alp,\bet,n)$ (\ref{eq4.2.1}) for $\alp=B$, $\bet=A$ as a function of the number of Floquet cycles $n$ for an hypothetical P-violating matrix element of $\Hw/h=1\,$kHz. In this case, the overall decay determined by $\xi$ is weak enough such that the first Floquet beat shows a clear P-violating change under a reversal of the fields' handedness. In Fig.\ \ref{fig5.7} also the complete time evolution of the probabilities $|\rbrak{\alp}\unU(t,0)\rket{\bet}|^2$ (cf.\ (\ref{eq2.4})) in between the Floquet points is shown. Fig.\ \ref{fig5.8} illustrates that the oscillations may occur for only one chirality (cf.\ (\ref{eq5.7.1}), (\ref{eq5.7.2})).

In summary, with the arrangement investigated so far it will not be possible to actually observe P-violating {\it oscillations} of the occupation numbers, which here, due to the reality of $\lamtil_1$, would only show up for one handedness of the fields. On the contrary, most of the atoms will have decayed long before one such oscillation is completed. However, as we will show in the following, it should be possible nevertheless to observe large P-violating effects.
%
%
%
%
%
%
%
%
%
%
%
\begin{figure}[tb]
\begin{center}
\setlength{\unitlength}{0.7mm}

\begin{picture}(190,110)(0,-15)

\put(0,0){\epsfig{file=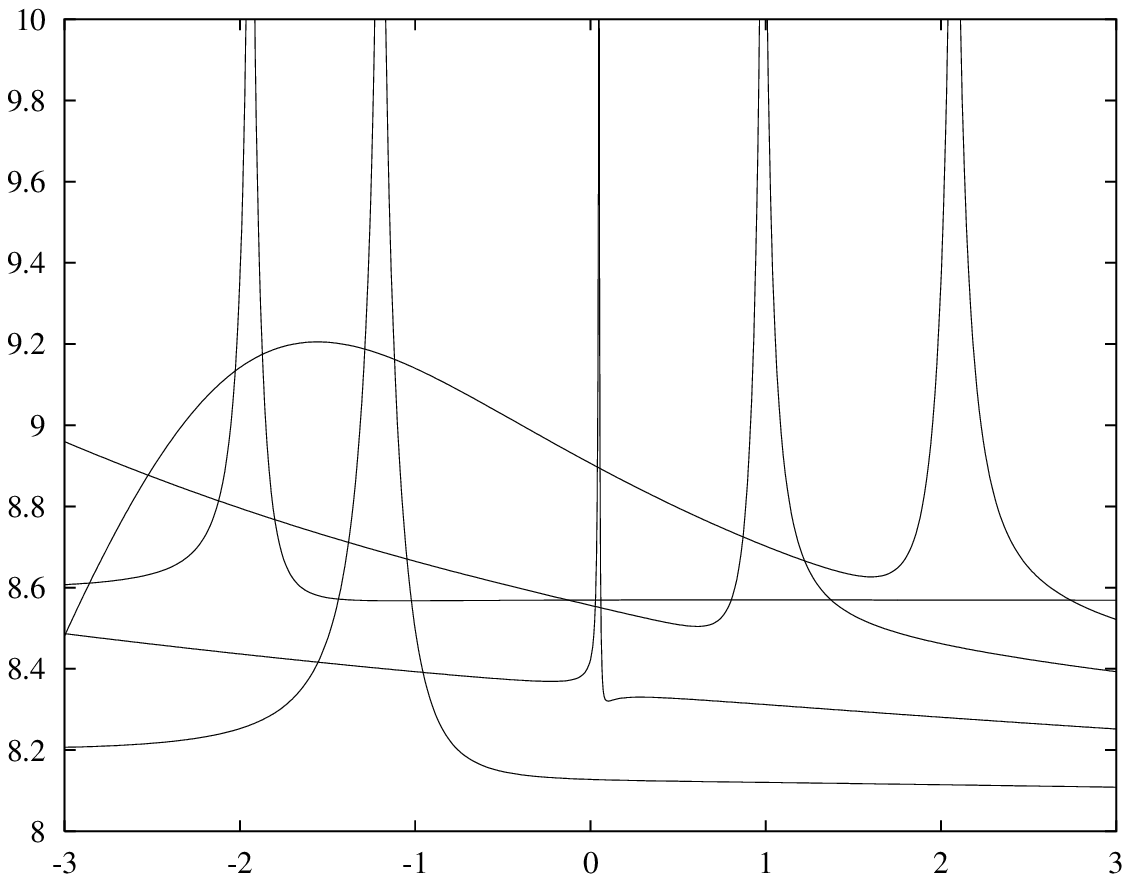,width=180\unitlength,height=100\unitlength}}


\put( 88,-10){$d\Del\cE^\iKn/h$ [kHz]}

\begin{rotate}{\Rotwinkel}
\put(45,-5){$\lg\Nbar_0$}
\end{rotate}

\put(35,90){$\ssty n\,=\,2$}
\put(60,90){$\ssty 6$}
\put(92,90){$\ssty 10$}
\put(115,90){$\ssty 14$}
\put(143,90){$\ssty 18$}

\end{picture}
\caption[ ]{\label{fig5.3}\fns The decadic logarithm of the minimum number $\Nbar_0$ (\ref{eq4.9.4}) of initial atoms for $\rket{\alp}=\rket{\bet}=\rket{A}$, $n=2,6,10,14,18$, for the same configuration of electric and magnetic fields as in Fig.\ \ref{fig5.2}.
The scale of the horizontal axis is stretched and denotes the difference $d\Del\cE^\iKn$ from the shifted central values $d\cE^\iKn_c$ (\ref{eq5.10}) as given in Table \ref{tab5.2} for each $n$. Contributions of order $\Hw^2$ have been neglected.}
\end{center}
\end{figure}

%
%


\subsection{Results for $A(\alp,\bet,n)$ and ${\overline{N}}_0$}
\label{sec5.2}
In this section we give the results of an optimization procedure for the asymmetry $A(\alp,\bet,n)$ (\ref{eq4.9}) and the minimum initial number $\Nbar_0$ (\ref{eq4.9.4}) of atoms required for a 1-$\sig$-effect. For this we introduce a
``premixing'' of the states between $t=0$ and $t=t_0$ by applying electric and magnetic fields $\cE^\iKn\evd, \cB^\iKn\evd$ (\ref{eq2.20}).

We considered an evolution of $n\le 20$ Floquet cycles, which corresponds to a time of approximately $60\,\mu$s (cf.\ (\ref{eq5.3}), Table \ref{tab5.1}). Within this period, $98\%$ of the initial number of states have decayed. However, for $n\le20$, the modulus of the P-violating arguments of the hyperbolic sine and cosine in (\ref{eq3.27})\,f.\ is smaller than $0.05$. 
In Fig.\ \ref{fig5.2} we show the asymmetry $A(\alp,\bet,n)$ (\ref{eq4.9}) for $\alp=\bet=A$, $1\le n\le20$, $\{\cEv^\iKsig,\Bv^\iKsig,\Del t^\iKsig; \sig=1,2,3\}$ as given in Table \ref{tab5.1}, choosing $\cB^\iKn=-\Del/(\mu_A-\mu_B)=1.4762\,$G, i.e.\ $\Del+(\mu_A-\mu_B)\cB^\iKn=0$, as a function of $d\cE^\iKn$. In computing $A(\alp,\bet,n)$ we neglected for the first contributions of order $\Hw^2$. We find an optimal asymmetry for $\rket{\alp}=\rket{\bet}=\rket{A}$, $n=10$, and $\cE^\iKn\simeq 0.28\,$V/cm of 
$A(\alp,\bet,n)=-0.84$.
%
%
%
%
%
%
%
%
%
%
%
\small

\begin{table}[tb]
\renewcommand{\arraystretch}{1.5}
\renewcommand{\tabcolsep}{0.2cm}
\begin{center}

\begin{tabular}{c|cccccccccc}
\hline
\hline 
 $n$ &1&2&3&4&5&6&7&8&9&10\\
 $d\cE_c^\iKn/h$ [kHz]   & $67.7$ & $61.2$ & $53.7$ & $45.1$ & $36.1$ &
			   $27.4$ & $19.0$ & $11.9$ &  $6.1$ &  $1.1$\\
\hline
 $n$ &11&12&13&14&15&16&17&18&19&20\\
 $d\cE_c^\iKn/h$ [kHz]   &  $-3.0$ &  $-6.4$ &  $-9.1$ & $-11.5$ & $-13.5$ & 
			   $-15.2$ & $-16.7$ & $-18.0$ & $-19.1$ & $-20.0$\\
\hline 
\hline
\end{tabular}
\caption[ ]{\label{tab5.2}\fns The central values $d\cE^\iKn_c$ (cf.\ (\ref{eq5.10})) of the resonances in Fig.\ \ref{fig5.2} for $n=1,...,20$.}
\end{center}
\renewcommand{\arraystretch}{1.0}
\end{table}
\normalsize

%
%
%
%
%
%
%
%
%
%
%
%
%
\begin{figure}[tb]
\begin{center}
\setlength{\unitlength}{0.7mm}

\begin{picture}(190,110)(0,-15)

\put(0,0){\epsfig{file=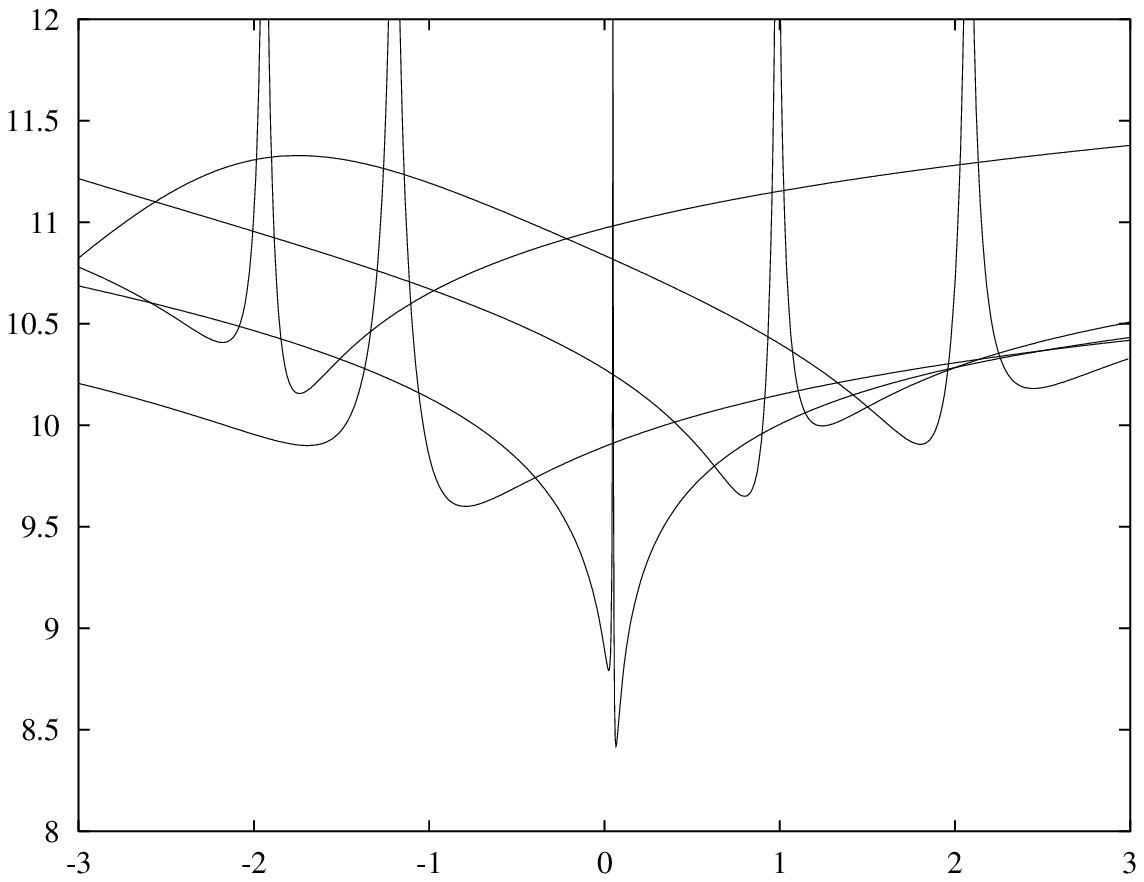,width=180\unitlength,height=100\unitlength}}


\put( 80,-10){$d\Del\cE^\iKn/h$ [kHz]}

\begin{rotate}{\Rotwinkel}
\put(30,-5){$\lg(\Nbar_0/|A(\alp,\bet,n)|)$}
\end{rotate}

\put(33,90){$\ssty n\,=\,2$}
\put(60,90){$\ssty 6$}
\put(92,90){$\ssty 10$}
\put(115,90){$\ssty 14$}
\put(143,90){$\ssty 18$}

\end{picture}
\caption[ ]{\label{fig5.4}\fns The decadic logarithm of the fraction $\Nbar_0/|A(\alp,\bet,n)|$ of initial atoms for $\rket{\alp}=\rket{\bet}=\rket{A}$, $n=2,6,10,14,18$, for the same configuration of electric and magnetic fields as in Fig.\ \ref{fig5.2}.
The scale of the horizontal axis is stretched and denotes the difference $d\Del\cE^\iKn$ from the shifted central values $d\cE^\iKn_c$ (\ref{eq5.10}) as given in Table \ref{tab5.2} for each $n$. Contributions of order $\Hw^2$ have been neglected.}
\end{center}
\end{figure}

%
%
In Fig.\ \ref{fig5.3} we show the decadic logarithm of the minimum number $\Nbar_0$ (\ref{eq4.9.4}) of initial atoms for $\rket{\alp}=\rket{\bet}=\rket{A}$, $n=2,6,10,14,18$, for the same configuration of electric and magnetic fields as in Fig.\ \ref{fig5.2}.
Note however, that for an appropriate resolution the scale of the horizontal axis is stretched and denotes the difference $d\Del\cE^\iKn$ from the central values $d\cE^\iKn_c$, shifted by $(10-n)/4\,$kHz$\,h$: 
\be
\label{eq5.10}
  d\cE^\iKn = d\cE^\iKn_c + \ev(10-n)\mbox{kHz}\,h+d\Del\cE^\iKn,
\ee
as given in Tab.\ \ref{tab5.2} for each $n$. 
%
%
%
%
%
%
%
%
%
%
%
\begin{figure}[tb]
\begin{center}
\setlength{\unitlength}{0.7mm}

\begin{picture}(100,110)(0,-15)

\put(0,0){\epsfig{file=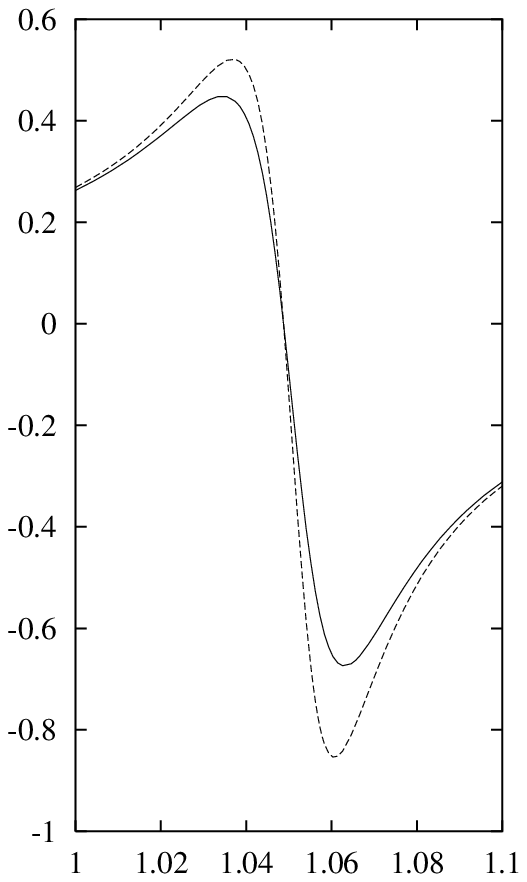,width=90\unitlength,height=100\unitlength}}


\put(-5,100){\bf (a)}
\put( 35,-10){$d\cE^\iKn/h$ [kHz]}

\begin{rotate}{\Rotwinkel}
\put(35,-5){$A(\alp,\bet,n)$}
\end{rotate}

\end{picture}
\hspace{0.5cm}
\begin{picture}(100,110)(0,-15)

\put(0,0){\epsfig{file=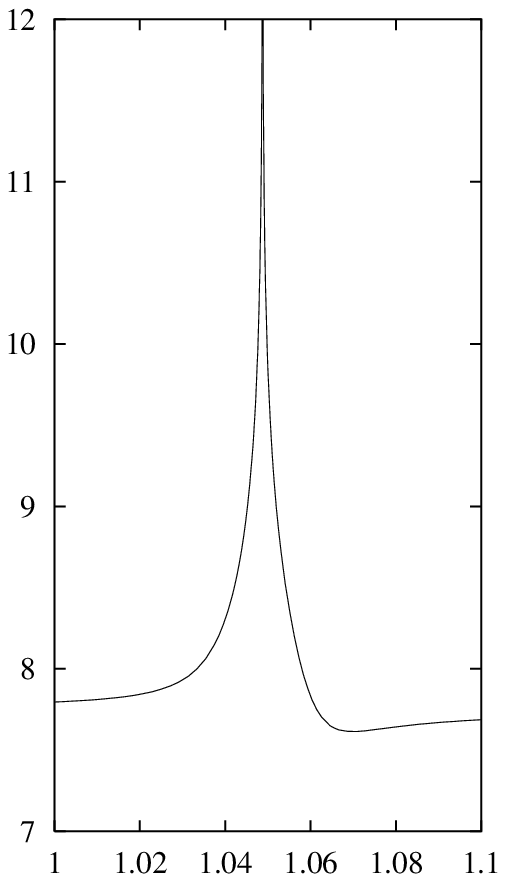,width=90\unitlength,height=100\unitlength}}


\put(-5,100){\bf (b)}
\put( 35,-10){$d\cE^\iKn/h$ [kHz]}

\begin{rotate}{\Rotwinkel}
\put(40,-5){$\lg\Nbar_0$}
\end{rotate}

\end{picture}
\vspace{0.5cm}\nopagebreak\newline
\begin{picture}(100,110)(0,-15)

\put(0,0){\epsfig{file=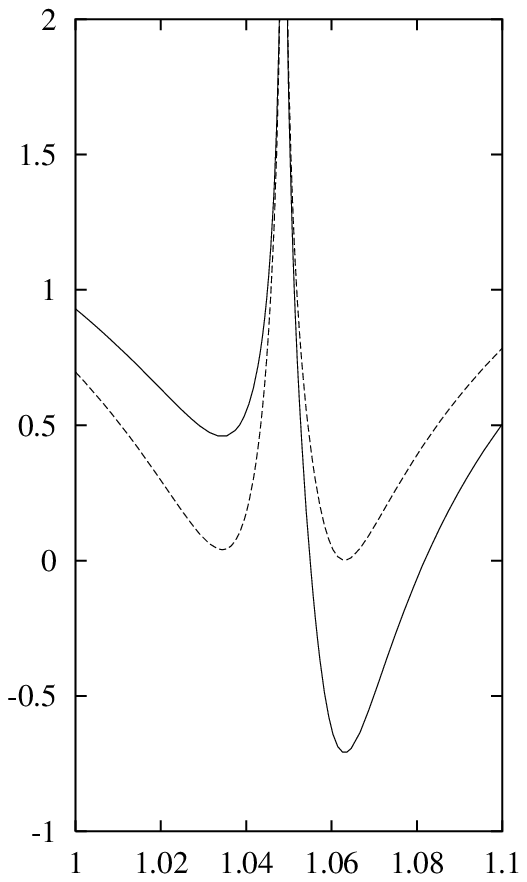,width=90\unitlength,height=100\unitlength}}


\put(-5,100){\bf (c)}
\put( 35,-10){$d\cE^\iKn/h$ [kHz]}

\begin{rotate}{\Rotwinkel}
\put(30,-5){$\lg(\Nbar_0 w_{\rI,\rII}(\alp,\bet,n))$}
\end{rotate}

\put(63,25){$\rI$}
\put(63,45){$\rII$}

\end{picture}
\hspace{0.5cm}
\begin{picture}(100,110)(0,-15)

\put(0,0){\epsfig{file=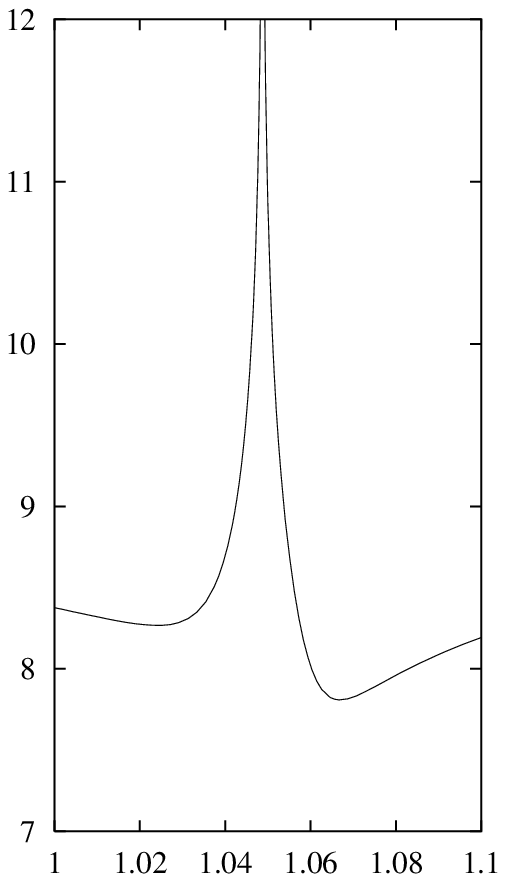,width=90\unitlength,height=100\unitlength}}


\put(-5,100){\bf (d)}
\put( 35,-10){$d\cE^\iKn/h$ [kHz]}

\begin{rotate}{\Rotwinkel}
\put(30,-5){$\lg(\Nbar_0/|A(\alp,\bet,n)|)$}
\end{rotate}

\end{picture}
\caption[ ]{\label{fig5.5}\fns (a) $A(\alp,\bet,n)$, (b) $\lg\Nbar_0$, (c) $\lg(\Nbar_0 w_{\rI,\rII}(\alp,\bet,n))$ and (d) $\lg(\Nbar_0/|A(\alp,\bet,n)|)$ for $n=10$, $\alp=\bet=A$, and the other pa\-ra\-meters chosen as in Figs.\ \ref{fig5.2}--\ref{fig5.4}. In (a) the dashed line represents the approximation, where terms of order $\Hw^2$ are neglected (cf.\ Fig.\ \ref{fig5.2}), the solid line is the exact asymmetry (\ref{eq4.9}). In Fig.\ (c) the solid line represents $\lg(\Nbar_0w_\rI)$, the dashed line $\lg(\Nbar_0w_\rII)$. Thus (c) shows how the occupation number changes under a reversal of the fields' handedness.}
\end{center}
\end{figure}

%
%
%
%
%
%
%
%
%
%
%
%
%
\begin{figure}[tb]
\begin{center}
\setlength{\unitlength}{0.7mm}

\begin{picture}(190,110)(0,-15)

\put(0,0){\epsfig{file=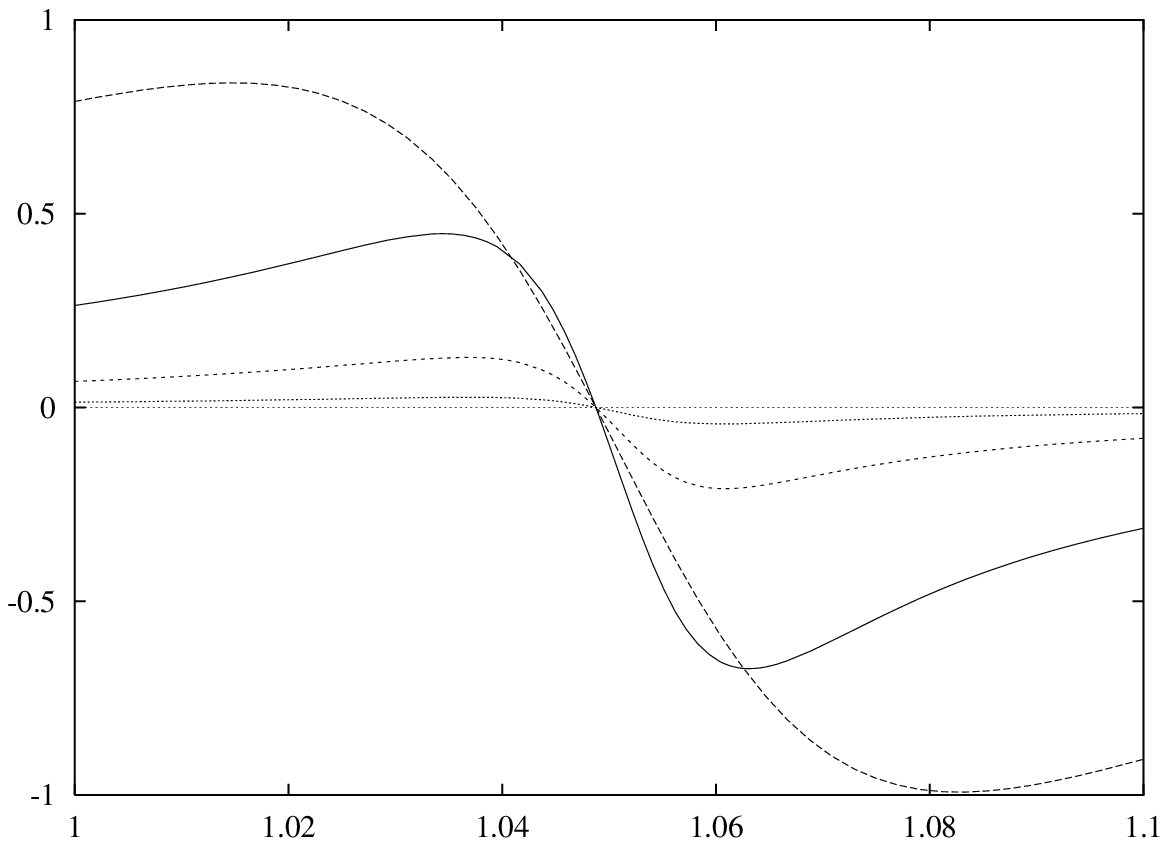,width=180\unitlength,height=100\unitlength}}


\put( 80,-10){$d\cE^\iKn/h$ [kHz]}

\begin{rotate}{\Rotwinkel}
\put(40,-5){$A(\alp,\bet,n)$}
\end{rotate}

\put(28,90){$\ssty \Hw/h\,=\,8\,\mathrm{Hz}$}
\put(28,67){$\ssty 2\,\mathrm{Hz}$}
\put(28,57){$\ssty 0.5\,\mathrm{Hz}$}
\thinlines
\put(28,45){$\ssty 0.1\,\mathrm{Hz}$}
\put(40,47){\vector(2,1){10}}

\end{picture}
\caption[ ]{\label{fig5.9}\fns The asymmetry $A(\alp,\bet,n)$ (\ref{eq4.9}) for $\alp=\bet=A$, $n=10$, $\{\cEv^\iKsig,\Bv^\iKsig,\Del t^\iKsig; \sig=1,2,3\}$ as given in Tab.\ \ref{tab5.1}, $\Del+(\mu_A-\mu_B)\cB^\iKn=0$, as a function of $d\cE^\iKn$ for different values of the P-violating matrix element $\Hw$.}
\end{center}
\end{figure}

%
%

In Fig.\ \ref{fig5.4} the decadic logarithm of the fraction $\Nbar_0/|A(\alp,\bet,n)|$ is plotted for the same parameters as in Fig.\ \ref{fig5.3}, again as a function of $\Del\cE^\iKn$. We find that, although the number of atoms $\Nbar_0$ increases when $n$ becomes larger, as can be expected from (\ref{eq4.9.4}) which is proportional to the factor (cf.\ (\ref{eq5.2}), (\ref{eq5.8}))
\be
\label{eq5.11}
  |\xi|^{-2n} = \exp\{0.22\,n\} = 10^{0.094\,n},
\ee
the observable $A(\alp,\bet,n)$ at $n=10$ is still the optimal one in the following sense: Although the asymmetry at larger values of $n$ may have a higher statistical significance indicated by a smaller value of $\Nbar_0$, the smaller value of the observable might be dominated by systematic experimental errors. In this sense, the fraction of $\Nbar_0$ divided by $A(\alp,\bet,n)$ may provide a useful measure for an optimal compromise when taking into account different sources of errors.   

%
%
%
%
In Figs.\ \ref{fig5.5}a--d we plot $A(\alp,\bet,n)$ (\ref{eq4.9}), $\lg\Nbar_0$, $\lg(\Nbar_0 w_{\rI,\rII}(\alp,\bet,n))$, $\lg(\Nbar_0/|A(\alp,\bet,n)|)$ for $n=10$, $\alp=\bet=A$, and the other pa\-ra\-meters chosen as in Figs.\ \ref{fig5.2}--\ref{fig5.4}. From Fig.\ \ref{fig5.5}b we can see that $d\cE^\iKn/h=1.07\,$kHz ($\cE^\iKn=0.28\,$V/cm) gives the minimal value for 
the initial number of atoms required for a 1-$\sig$-effect:
\be
\label{eq5.12}
  \Nbar_0 = 4\cdot10^7.
\ee
There we have an asymmetry of
\be
\label{eq5.9}
A(\alp,\bet,10)=-0.7.
\ee
Fig.\ \ref{fig5.5}c shows that, starting with $4\cdot10^8$ atoms, after $n=10$ steps, one should observe either about $10$ or only $2$ atoms, depending on the handedness of the field configuration. Fig.\ \ref{fig5.5}d shows that $\lg(\Nbar_0/|A(\alp,\bet,n)|)$ is also almost at its minimum.

In Fig.\ \ref{fig5.9} we plot the asymmetry $A(\alp,\bet,n)$ (\ref{eq4.9}) for $\alp=\bet=A$, $n=10$, and the other parameters as in Fig.\ \ref{fig5.5}a, for different values of the P-violating matrix element $\Hw$. This illustrates, that $\Hw$ can be determined from the magnitude of the asymmetry and the positions of its extrema. 
We point out that also the sign of $\Hw$ may be determined from the asymmetry, since it changes sign under $\Hw\to-\Hw$.

We also varied the magnetic field $\cB^\iKn$ and found that slightly away from the Zeeman-level degeneracy $\Del+(\mu_A-\mu_B)\cB^\iKn=0$ one can even reach a $100\%$-effect! 
%
%
%
%
%
%
However, the number of atoms required to observe a 1-$\sig$-effect is of the same order of magnitude as above. Thus we will not investigate this further here. But it shows that the occurence of large P-violating asymmetries is not necessarily connected to the P-even degeneracy of the Floquet eigenvalues. 

To summarize, our results of this section show, that for a choice of parameters as given in Tab.\ \ref{tab5.1} large asymmetries may be reached easily within a few Floquet cycles, for premixing fields within an accessible range. 
\newpage

  

%
%
%
%
%
%
\section{Conclusions}
\label{secConcl}
\setcounter{equation}{0}
In this article we have investigated parity violating effects in atomic dysprosium where one has a near degeneracy of two levels of opposite parity. We considered the time evolution of this 2-level-system in external electric and magnetic fields having a periodic structure in time with period $T$. The object to study was the Floquet matrix which describes this periodic time evolution. We found that for unstable states the eigenvalues of the Floquet matrix which give the eigenfrequencies of the system can have contributions proportional to the square root of the P-violating weak interaction matrix element $\Hw$. This leads to beat frequencies proportional to $\sqrt{\Hw}$. For dysprosium in the simple arrangements of external fields which we considered we found, however, that it is difficult to observe such beat effects since the states decay too fast. On the other hand, we found that very large P-violating asymmetries are easily obtained after a few Floquet cycles. This is achieved by tuning carefully the initial electric and magnetic fields in which the system evolves before entering the Floquet cycles. The electric and magnetic field, the time constants etc.\ necessary for this are all in an experimentally accessible range. The number of atoms necessary to obtain a statistically significant signal is generally of order $10^{7}$ to $10^8$ if $|\Hw|/h\simeq2\,$Hz. Once having measured the asymmetry one may determine both the modulus and the sign of $\Hw$. It is clear that our considerations can easily be extended to other periodic electric and magnetic field arrangements and other 2-level atomic systems. We hope that these ideas will stimulate experimental work in this direction.

\ \vspace{1cm}\\
{\bf Acknowledgements:} The authors are grateful to M.\ Zolotorev and D.\ Budker for discussions and correspondence which led to these investigations. They would also like to thank T.\ Brunne for useful discussions and suggestions.

\newpage
\begin{appendix}
\section*{Appendix}
%
%
%
%
%
%
\renewcommand{\thesection}{\Alph{section}}
\setcounter{section}{0}
%
%
\section{Explicit form of the matrix $\unV$}
\label{appA}
\renewcommand{\theequation}{\Alph{section}.\arabic{equation}}
\setcounter{equation}{0}
In this appendix we give the explicit form of the matrix $\unV$ (cf.\ (\ref{eq3.9})), which rotates the matrices $\unU^\iKsig$, $\sig=1,...,\sigmax$, such that the trace (\ref{eq3.9}) looses any $\Hw$-odd contributions if (\ref{eq3.9.1}) is satisfied.

%
%
%
%
%
%
%
%
%
%
%
%
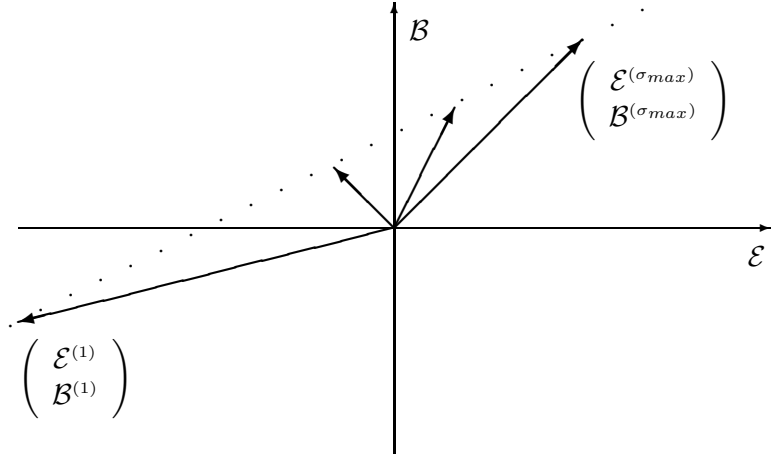
\begin{figure}[tb]
\begin{center}
\setlength{\unitlength}{1mm}
\begin{picture}(100,70)(0,0)
\small
\put( 0,35){\vector(1, 0){100}}
\put(50, 5){\vector(0, 1){60}}
\thicklines
\put(50,35){\vector(1, 1){25}}
\put(50,35){\vector(1, 2){8}}
\put(50,35){\vector(-1,1){8}}
\put(50,35){\vector(-4,-1){50}}
\thinlines
\multiput(-1,22)(4,2){22}{\circle*{0.05}}
\put(97,30){$\cE$}
\put(52,60){$\cB$}
\put( 0,14){$\colvecz{\cE^\iKe}{\cB^\iKe}$}
\put(74,51){$\colvecz{\cE^\iKvar{\sigmax}}{\cB^\iKvar{\sigmax}}$}
\end{picture}
\caption[ ]{\label{figA.1}\fns Sketch of the vectors $(\cE^\iKsig,\cB^\iKsig)$ for different $\sig$, chosen such that they lie on a straight line in the $\cE$--$\cB$-plane. For such configurations the Floquet eigenvalues have no P-violating contributions.}
\end{center}
\end{figure}
%

%
%
We start with a set of electric and magnetic field strengths' components $\{\cE^\iKsig,\cB^\iKsig;\,\sig=1,...,\sigmax\}$ lying on a straight line in the $\cE$-$\cB$-plane (cf.\ (\ref{eq3.9.1}) and Fig.\ \ref{figA.1}). Of course, $v_{\cE,\cB}, w_{\cE,\cB}$ and $\zeta^\iKsig$ in (\ref{eq3.9.1}) are real. To have a nontrivial field configuration we require
\be
\label{eqA.1}
  w_\cE^2+w_\cB^2 \not= 0.
\ee
The corresponding vectors $\hv^\iKsig$ ((\ref{eq2.28})--(\ref{eq2.30})) are then
\bea
\label{eqA.1.1}
  &\hv^\iKsig 
    = \colvecd{0}{-\Hw}{-\eh\Del-\iv(\Gam_A-\Gam_B)}
      +\vvtil+\wvtil\,\zeta^\iKsig,&\\
\label{eqA.1.2}
  &\vvtil 
    = \colvecd{d\,v_\cE}{0}{-\eh(\mu_A-\mu_B)v_\cB},\qquad
   \wvtil 
    = \colvecd{d\,w_\cE}{0}{-\eh(\mu_A-\mu_B)w_\cB}.& 
\eea
Without loss of generality we may assume
\be
\label{eqA.1.3}
  \wvtil\cdot\wvtil = 1,\qquad \vvtil\cdot\wvtil = 0.
\ee
The sought matrix $\unV$ may be composed out of two separate transformations,
\be
\label{eqA.2}
  \unV = \unV_2\,\unV_1.
\ee
We choose the first rotation to be one about the 2-axis, such that $\wvtil$ becomes parallel to the 1-axis. The second one is chosen to be a rotation about the 1-axis. This will bring all vectors (\ref{eqA.1.1}) into the 1-3-plane. Our ansatz thus reads:
\bea
\label{eqA.3}
  \unV_1 
  &=& \exp\lk\{\ih\alp\,\sig_2\rk\}
   =  \lk(\ba{cc} \cos(\alp/2) & \sin(\alp/2) \\ -\sin(\alp/2) & \cos(\alp/2)\ea\rk),\\
\label{eqA.4}
  \unV_2
  &=& \exp\lk\{\ih\bet\,\sig_1\rk\}
   =  \lk(\ba{cc} \cos(\bet/2) & i\sin(\bet/2) \\ i\sin(\bet/2) & \cos(\bet/2)\ea\rk).
\eea
Choosing $\alp$ such that
\be
\label{eqA.5}
  \cos\alp = \wtil_1,\qquad
  \sin\alp = -\wtil_3
\ee
we find:
\be
\label{eqA.6}
  \unV_1\unheff^\iKsig\unV_1^{-1} = \sum_{i=0}^3\,h_i^{'\iKsig}\sig_i,
\ee
with
\bea
\label{eqA.7}
  h_0^{'\iKsig} 
  &=& h_0^\iKsig,\\
\label{eqA.8}
  h_1^{'\iKsig} 
  &=& h_1^\iKsig\cos\alp-h_3^\iKsig\sin\alp\nonumber\\
  &=& -\wtil_3\lk[\eh\Del+\iv(\Gam_A-\Gam_B)\rk] +\zeta^\iKsig,\\
\label{eqA.9}
  h_2^{'\iKsig} 
  &=& h_2^\iKsig = -\Hw,  \\
\label{eqA.10}
  h_3^{'\iKsig} 
  &=& h_3^\iKsig\cos\alp+h_1^\iKsig\sin\alp\nonumber\\
  &=& -\wtil_1\lk[\eh\Del+\iv(\Gam_A-\Gam_B)\rk] + (\vvtil\cross\wvtil)\cdot\evz
      =: h'_3.
\eea
Note that $h_2^{'\iKsig}$ and $h_3^{'\iKsig}$ are the same for all $\sig$.
The angle $\beta$ of the second rotation is chosen to satisfy:
\bea
\label{eqA.11}
  \cos\beta 
  &=& h_3'\lk[(h_3')^2+\Hw^2\rk]^{-1/2},\nonumber\\
  \sin\beta 
  &=& \Hw\lk[(h_3')^2+\Hw^2\rk]^{-1/2}.
\eea
Here we assume $h_3'\not=\pm i\Hw$ which will hold except for very special cases. It is clear that $\bet$ is in general complex, near zero and odd under the exchange $\Hw\to-\Hw$.
Using (\ref{eqA.2}), (\ref{eqA.4}) and (\ref{eqA.6})--(\ref{eqA.11}) we find 
\be
\label{eqA.14}
  \unV\unheff^\iKsig\unV^{-1} 
  = \unV_2\,\lk(\sum_{i=0}^{3}\,h_i^{'\iKsig}\sig_i\rk)\,\unV_2^{-1} 
  = \sum_{i=0}^3\,h_i^{''\iKsig}\sig_i,
\ee
with
\bea
\label{eqA.15}
  h_0^{''\iKsig} 
  &=& h_0^\iKsig,\\
\label{eqA.16}
  h_1^{''\iKsig} 
  &=& h_1^{'\iKsig},\\
\label{eqA.17}
  h_2^{''\iKsig} 
  &=& 0,\\
\label{eqA.18}
  h_3^{''\iKsig} 
  &=& \sqrt{(h_3')^2+\Hw^2}.
\eea
We see from (\ref{eqA.15})--(\ref{eqA.18}) that the $h_j^{''\iKsig}$, $j=0,...,3$, are all even functions of $\Hw$. Inserting $h_j^{''\iKsig}$ instead of $h_j^\iKsig$ in (\ref{eq2.26})--(\ref{eq2.35}) we see that also the corresponding Floquet matrix $\unU''(t_0+T,t_0)$ is an even function of $\Hw$. The same is then true for its trace which is the same as the trace of the original Floquet matrix according to (\ref{eq3.9}):
\be
\label{eqA.19}
  \Tr\unU(t_0+T,t_0) = \Tr\unU''(t_0+T,t_0) = \mbox{even function of $\Hw$}.
\ee
Using also (\ref{eq3.5}) and (\ref{eq2.8})--(\ref{eq2.10}) we prove our assertions made in Section \ref{sec3}, that if the points ($\cE^\iKsig,\cB^\iKsig$) lie on a straight line, the Floquet eigenvalues are even functions of $\Hw$ and thus have no P-odd contributions.

%
%
%
%
%
%
%
%
\section{The case of stable states, $\Gam_A=\Gam_B=0$}
\label{appB}
\renewcommand{\theequation}{\Alph{section}.\arabic{equation}}
\setcounter{equation}{0}
In this appendix we show explicitly for the case of a time-evolution with $\sigmax=3$ sections of constant fields within one period of oscillation, that for stable states, i.e. $\Gam_A=\Gam_B=0$, conditions (\ref{eq3.29}) and (\ref{eq3.30}) may not be fulfilled simultaneously. As stated in section \ref{sec3}, $\lamtil_1$ (\ref{eq3.16}) will vanish in any case where $\lamtil_0^2$ (\ref{eq3.14}) is equal to 1.

The reason for this becomes clear from the following general argument: For stable states the Floquet matrix $\unU$ is unitary. A unitary matrix $\unU$ may always be written as $\unU=\exp\{-i\unH T\}$, where $\unH$ is a hermitean matrix. For the case that $\unU$ and $\unH$ are (2$\times$2)-matrices, the eigenvalues of $\unH$ are given as
\be
\label{eqB.0.1}
  \ome_\pm = \eh\Tr\unH\pm\sqrt{(\eh\Tr\unH)^2-\det\unH}.
\ee
Let the series expansion of matrix elements of $\unH^\iKn$ w.r.t.\ the small parameter $\del:=\Hw/\Lambda$, where $\Lambda$ is some constant with the dimensions of $\Hw$, be $H_{ij}=H_{ij}^\iKn+\del H_{ij}^\iKe+\cO(\del^2)$. Then the expansion of the eigenvalues (\ref{eqB.0.1}) reads
\bea
\label{eqB.0.2}
  \ome_\pm 
  &=& \eh\lk(H_{11}^\iKn+H_{22}^\iKn\rk)
      \pm\eh\Big\{\lk(H_{11}^\iKn-H_{22}^\iKn\rk)^2
      +4H_{12}^\iKn H_{21}^\iKn
\nonumber\\
  &&\ +\ 2\del\lk[\lk(H_{11}^\iKn-H_{22}^\iKn\rk)
                  \lk(H_{11}^\iKe-H_{22}^\iKe\rk)
      +2(H_{12}^\iKe H_{21}^\iKn+H_{12}^\iKn H_{21}^\iKe)\rk]\Big\}^{1/2}
\nonumber\\  
  &&\ +\ \cO(\del).
\eea
Now, if the eigenvalues are degenerate to the order $\del^0$, we may diagonalize $\unH$ such that $H_{11}^\iKn=H_{22}^\iKn$ and $H_{12}^\iKn=H_{21}^\iKn=0$. Then it becomes immediately clear from (\ref{eqB.0.2}), that the radicand is zero up to order $\del$. Hence, no contributions to $\ome_\pm$ proportional to $\sqrt{\del}$ are possible. The same is then true for the eigenvalues of $\unU$. Note also that, from the physical point of view, such contributions would make $\ome_\pm$ complex for one handedness of the electric and magnetic fields, which can not be true since $\unH$ is hermitean.

We now want to prove that for $\Gam_A=\Gam_B=0$ the expressions (\ref{eq3.14}) and (\ref{eq3.15}) have this property for any field configuration. 
In order to do this we will first verify a lemma for the trace of the product of two unitary matrices $\unU^\iKsig$ of the form (\ref{eq2.31}). Let us first introduce the notation
\bea
\label{eqB.1}
  s^\iKsig &:=& \sin(\eta^\iKsig\Del t^\iKsig),\nonumber\\
  c^\iKsig &:=& \cos(\eta^\iKsig\Del t^\iKsig),\nonumber\\
  s^\iKvar{\pm} 
  &:=& \sin\lk(\frac{\eta^\iKz\Del t^\iKz\pm\eta^\iKe\Del t^\iKe}{2}\rk),
  \nonumber\\
  c^\iKvar{\pm} 
  &:=& \cos\lk(\frac{\eta^\iKz\Del t^\iKz\pm\eta^\iKe\Del t^\iKe}{2}\rk),
  \nonumber\\
  \hvh^\iKsig &:=& \hv^\iKsig/\eta^\iKsig,\nonumber\\
  c^\iKvar{\sig\tau} &:=& \hvh^\iKsig\cdot\hvh^\iKtau.
\eea
Note that for stable states, $\eta^\iKsig$ (\ref{eq2.34}) is the length of the real vector $\hv^\iKsig$ and thus $\hvh^\iKsig$ is a unit vector and $c^\iKvar{\sig\tau}$ the cosine of the angle between two such vectors.
We will prove now the

{\sl Lemma:} Given the half of the trace of the product of two unitary matrices $\unU^\iKe,\unU^\iKz$, with determinants $\det\unU^\iKsig=1,\sig=1,2$, i.e.\ $h_0^\iKsig=0$ (cf.\ (\ref{eq2.26})--(\ref{eq2.35})),
\bea
\label{eqB.2}
  T_{21} 
  &=& \eh\Tr\lk(\unU^\iKz\unU^\iKe\rk)\nonumber\\
  &=& c^\iKz c^\iKe - c^\iKvar{21} s^\iKz s^\iKe, 
\eea
its modulus may then and only then equal one,
\be
\label{eqB.3}
  |T_{21}| = 1,
\ee
if either
\be
\label{eqB.4}
  |c^\iKz| = |c^\iKe| = 1
\ee
or
\be
\label{eqB.5}
  |c^\iKvar{21}| = 1,
\ee
i.e.\ if both matrices equal plus or minus the unit matrix or $\hv^\iKe$ and $\hv^\iKz$ are parallel to each other.

For the proof one first calculates
\bea
\label{eqB.6}
  1-c^\iKz c^\iKe &=& {s^\iKvar{+}}^2+{s^\iKvar{-}}^2,\\
\label{eqB.7}
  1+c^\iKz c^\iKe &=& {c^\iKvar{+}}^2+{c^\iKvar{-}}^2,\\
\label{eqB.8}
    s^\iKz s^\iKe &=& {s^\iKvar{+}}^2-{s^\iKvar{-}}^2\\
\label{eqB.9}
                  &=& {c^\iKvar{-}}^2-{c^\iKvar{+}}^2.
\eea
From this we obtain the equivalences:
\bea
\label{eqB.10}
  T_{21} =  1\quad&\Leftrightarrow&\quad
  1-c^\iKz c^\iKe = c^\iKvar{21}s^\iKz s^\iKe,\\
\label{eqB.11}
  &\Leftrightarrow&\quad
  {s^\iKvar{+}}^2+{s^\iKvar{-}}^2 = c^\iKvar{21}\lk(
  {s^\iKvar{+}}^2-{s^\iKvar{-}}^2\rk),\\
\label{eqB.12}
  T_{21} = -1\quad&\Leftrightarrow&\quad
  1+c^\iKz c^\iKe = -c^\iKvar{21}s^\iKz s^\iKe,\\
\label{eqB.13}
  &\Leftrightarrow&\quad
  {c^\iKvar{+}}^2+{c^\iKvar{-}}^2 = c^\iKvar{21}\lk(
  {c^\iKvar{+}}^2-{c^\iKvar{-}}^2\rk).
\eea
In the case that $|c^\iKvar{21}| < 1$, it follows immediately from
both (\ref{eqB.11}) and (\ref{eqB.13}) --- since $0\le {s^\iKvar{\pm}}^2\le1$ and $0\le {c^\iKvar{\pm}}^2\le1$ ---, together with (\ref{eqB.8}), (\ref{eqB.9}), that
\be
\label{eqB.14}
  s^\iKz s^\iKe = 0
\ee
and thus from (\ref{eqB.10}) and (\ref{eqB.12}) that (\ref{eqB.4}) is necessary for (\ref{eqB.3}). For $c^\iKvar{21} = \pm1$ (\ref{eqB.5}) it follows from (\ref{eqB.11}) and (\ref{eqB.13}) that $T_{12}=1$ for $s^\iKvar{\mp}=0$ and $T_{21}=-1$ for $c^\iKvar{\mp}=0$. {\sl qed}.\\

By applying this Lemma to the trace of the product of a third matrix $\unU^\iKd$ with the product of the first two matrices, we will now prove the initial statement. 
In the follwing $\Hw$ will be set to zero if not otherwise stated. Then the (real) vectors $\hv^\iKsig$, $\sig=1,...,3$, all lie within the 1-3-plane (cf.\ (\ref{eq2.28})--(\ref{eq2.30})).
Consider the product 
\bea
\label{eqB.16}
  \unU^\iKvar{21} 
  &=& \unU^\iKz\unU^\iKe\nonumber\\
  &=& \lk(c^\iKz c^\iKe - c^\iKvar{21} s^\iKz s^\iKe\rk)\eins\nonumber\\
  & &\ -\ i\lk[\hvh^\iKe c^\iKz s^\iKe + \hvh^\iKz s^\iKz c^\iKe
               +\lk(\hvh^\iKz\cross\hvh^\iKe\rk)s^\iKz s^\iKe\rk]\cdot\sigv.
\eea
Defining
\be
\label{eqB.17}
  \hv^\iKvar{21} := \hvh^\iKe c^\iKz s^\iKe + \hvh^\iKz s^\iKz c^\iKe
                    +\lk(\hvh^\iKz\cross\hvh^\iKe\rk)s^\iKz s^\iKe
\ee
one finds (cf.\ (\ref{eqB.2}))
\be
\label{eqB.18}
  |\hv^\iKvar{21}|^2 = 1-T_{21}^2.
\ee
Therefore (\ref{eqB.16}) may be written as
\be
\label{eqB.19}
  \unU^\iKvar{21} = \eins \cos(|\hv^\iKvar{21}|)-i\,\hvh^\iKvar{21}\cdot\sigv\,
                          \sin(|\hv^\iKvar{21}|),
\ee
with $\hvh^\iKvar{21}=\hv^\iKvar{21}/|\hv^\iKvar{21}|$.

Let us now consider the trace (divided by 2)
\bea
\label{eqB.20}
  T_{321} 
  &=& \eh\Tr\lk(\unU^\iKd\unU^\iKz\unU^\iKe\rk)\nonumber\\
  &=& \eh\Tr\lk(\unU^\iKd\unU^\iKvar{21}\rk)
\eea
and investigate two cases: (a) $|c^\iKvar{21}|=1$, i.e.\ $s^\iKvar{21}=(1-{c^\iKvar{21}}^2)^{1/2}=0$, and (b) $|c^\iKvar{21}|<1$, i.e.\ $s^\iKvar{21}\not=0$.

In case (a) we have
\be
\label{eqB.21}
  \hv^\iKvar{21} = \hvh^\iKe\lk(c^\iKz s^\iKe + s^\iKz c^\iKe\rk).
\ee
From the above Lemma we therefore conclude that  
\be
\label{eqB.22}
  |T_{321}|=1
\ee
is only possible if either   
\be
\label{eqB.23}
  |\hvh^\iKd\cdot\hvh^\iKvar{21}|=1,
\ee
and thus all ``points'' $\hv^\iKsig$, $\sig=1,2,3$, lie on a straight line, or if  
\be
\label{eqB.24}
  |c^\iKd|=|\cos(|\hv^\iKvar{21}|)|=|T_{21}|=1
\ee
and thus $s^\iKd=0$. Note that (\ref{eqB.24}) may be fulfilled again due to the Lemma. 

In case (b) the vector $\hv^\iKvar{21}$ does not lie in the 1-3-plane which is spanned by the vectors $\hv^\iKz$ and $\hv^\iKe$ because of the cross product term in (\ref{eqB.17}). Then, again due to the above Lemma, (\ref{eqB.22}) requires (\ref{eqB.24}) to be fulfilled since (\ref{eqB.23}) is not possible
as $\hv^\iKd$ lies within the 1-3-plane (remember that $\Hw=0$). Then $s^\iKe=s^\iKz=0$ due to the Lemma, and $s^\iKd=0$ from (\ref{eqB.24}).

In summary, in any case where (\ref{eqB.22}) is fulfilled for stable states, i.e.\ $\lamtil_0^2=1$ (cf.\ (\ref{eq3.14}), (\ref{eq3.17})\,ff.), the $\Hw$-linear term $2\xi^2\lamtil_0\lamtil_1$, and thus $\vep$ (\ref{eq3.23}) vanishes as well, either by the arguments presented in Appendix \ref{appA} or by the sine $s^\iKd$ being zero (cf.\ (\ref{eq3.16})). {\sl qed.} 

\end{appendix}

\newpage
%
%
%
%
}
\end{document}